\documentclass[twocolumn,prb,aps,superscriptaddress,floatfix,nofootinbib]{revtex4-2}%
\usepackage{amsmath}
\usepackage[dvips]{graphicx}
\usepackage{upgreek}
\usepackage{dcolumn}
\usepackage{makeidx}
\usepackage{color}
\usepackage{float}
\usepackage{mathtools}
\usepackage{threeparttable}
\usepackage[linkcolor=blue,anchorcolor=black,citecolor=blue]{hyperref}%
\usepackage[T1]{fontenc} 
\newcommand{\ket}[1]{\left|#1\right>}
\newcommand{\bra}[1]{\left<#1\right|}

\begin{document}

\title{Balancing coherent and dissipative dynamics in a central-spin system}

\author{A.~Ricottone}
\email[]{alessandro.ricottone@mail.mcgill.ca}
\affiliation{Department of Physics, McGill University, Montr\'eal, Qu\'ebec H3A 2T8, Canada}

\author{Y.~N.~Fang}
\email{ynfang@csrc.ac.cn}
\affiliation{Beijing Computational Science Research Center, Beijing 100084, China}
\affiliation{Department of Physics, McGill University, Montr\'eal, Qu\'ebec H3A 2T8, Canada}
\affiliation{CAS Key Laboratory of Theoretical Physics, Institute of Theoretical Physics, Chinese Academy of Sciences, and University of the Chinese Academy of Sciences, Beijing 100190, China}
\affiliation{Synergetic Innovation Center of Quantum Information and Quantum Physics, University of Science and Technology of China, Hefei, Anhui 230026, China}

\author{W.~A.~Coish}
\email{coish@physics.mcgill.ca}
\affiliation{Department of Physics, McGill University, Montr\'eal, Qu\'ebec H3A 2T8, Canada}

\begin{abstract}
The average time required for an open quantum system to reach a steady state (the steady-state time) is generally determined through a competition of coherent and incoherent (dissipative) dynamics.  Here, we study this competition for a ubiquitous central-spin system, corresponding to a `central' spin-1/2 coherently coupled to ancilla spins and undergoing dissipative spin relaxation. The ancilla system can describe $N$ spins-1/2 or, equivalently, a single large spin of length $I=N/2$.  We find exact analytical expressions for the steady-state time in terms of the dissipation rate, resulting in a minimal (optimal) steady-state time at an optimal value of the dissipation rate, according to a universal curve.  Due to a collective-enhancement effect, the optimized steady-state time grows only logarithmically with increasing $N=2I$, demonstrating that the system size can be grown substantially with only a moderate cost in steady-state time.  This work has direct applications to the rapid initialization of spin qubits in quantum dots or bound to donor impurities, to dynamic nuclear-spin polarization protocols, and may provide key intuition for the benefits of error-correction protocols in quantum annealing.   
\end{abstract}

\maketitle

\section{Introduction}
The interplay of coherent and incoherent dynamics for a quantum system plays an important role, both in fundamental studies of thermalization and in the operation of a quantum processor (for example, a universal quantum computer or a specialized quantum simulator). While the state of a quantum computer can be manipulated via coherent dynamics to perform the actual computation, dissipation can be used to drive the system toward a certain target state, either for fast preparation of pure ancilla qubits or, e.g., to prepare the ground state of some more complicated local Hamiltonian.  For these applications, it may be important to tune experimental parameters in order to reach the target state as quickly as possible, or with the highest possible fidelity \cite{gordon2009cooling, alvarez2010zeno}. Reaching a particular target state may only be possible in some cases in the presence of a finite dissipation rate.  On the other hand, when the dissipation rate is too large, evolution toward a target state may be suppressed altogether (the quantum Zeno regime \cite{misra1977zeno, peres1980zeno}). The quantum Zeno regime has been studied in various settings, including circuit quantum electrodynamics (QED) \cite{gambetta2008quantum}, caviy QED \cite{bernu2008freezing, raimond2012quantum}, and in models with spin-bath interactions \cite{segal2007zeno}.  In this paper, we study the interplay of coherent dynamics and dissipation leading to a quantum Zeno regime for a particular central-spin model.

\begin{figure}
\centering
\includegraphics[width=0.96\columnwidth]{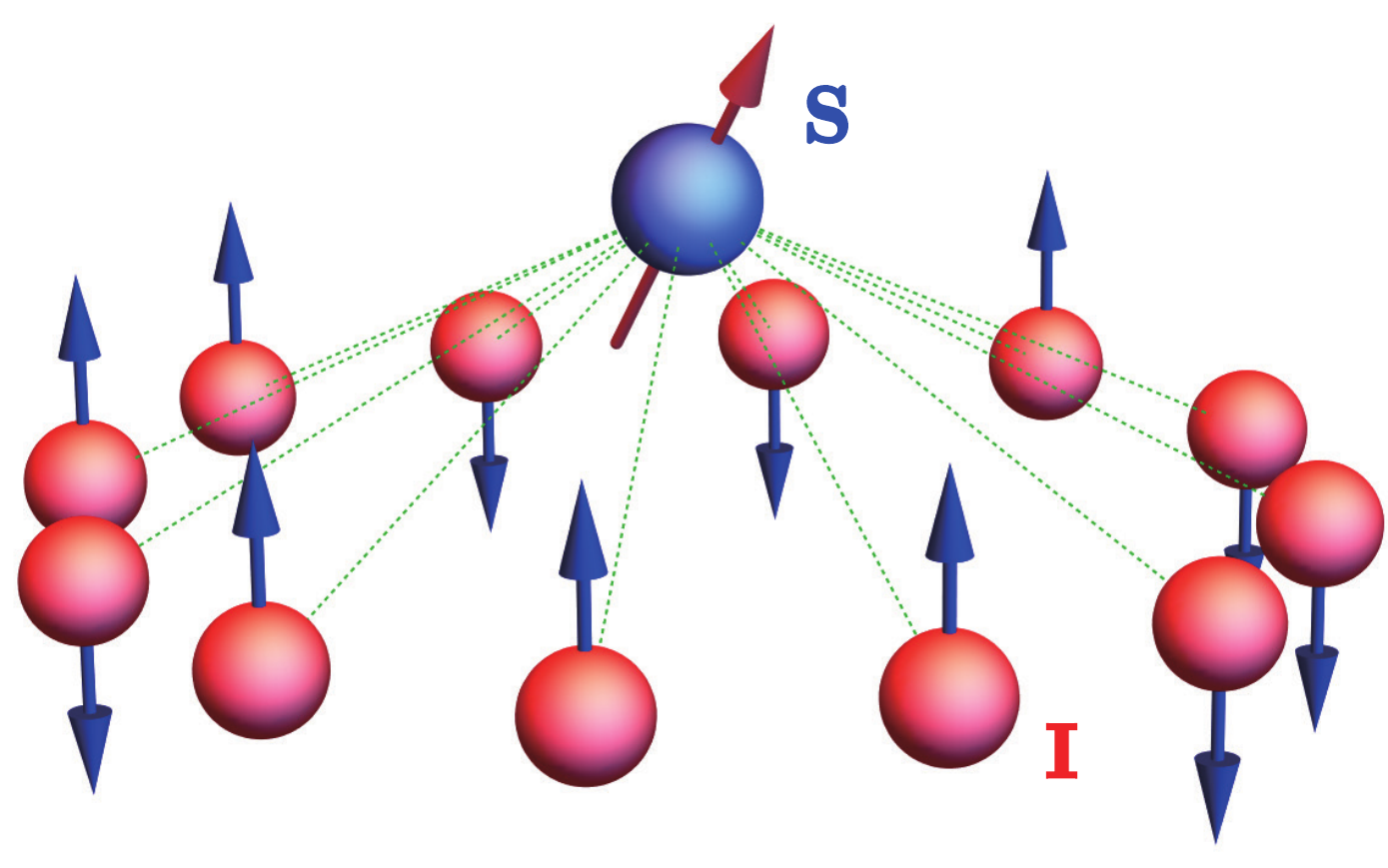}
\caption{(Color online)  A central-spin system, describing a spin $\mathbf{S}$ (S=1/2) coupled to $N$ ancilla spins $\mathbf{I}^l$ ($I^l=1/2$).  When the central spin is coupled equally to every ancilla, the total angular momentum of the ancillas is fixed (e.g., $I=N/2$), and the ancilla is described by a single large spin $\mathbf{I}=\sum_{l}\mathbf{I}^l$.}  \label{fig:css_sketch}
\end{figure}
A central-spin system (Fig.~\ref{fig:css_sketch}) may model a collection of coupled qubits or a single qubit exposed to a spin environment. In this paper, we consider a model where a spin-$1/2$, the central spin, interacts with an ancilla spin $I\ge 1/2$. The central spin-1/2 may describe an electron spin, a hole spin, or some other qubit.  When the central spin describes an electron, the ancilla spin may describe a second (exchange-coupled) electron in a double quantum dot \cite{watson2018programmable, zajac2018resonantly} ($I=1/2$), the nuclear spin of a donor impurity ($I\ge 1/2$) \cite{buch2013spin, franke2016quadrupolar, schenkel2006electrical, muhonen2018coherent}, or many nuclear spins locked in a total angular momentum eigenstate \cite{chekhovich2013nuclear, coish2009nuclear, kessler2012dissipative, fuchs2015thermal} ($I\gg 1/2$) [Fig.~\ref{fig:css_sketch}]. The central-spin system can be related to a class of dissipative quantum walks  \cite{rudner2010phase} and to the Tavis-Cummings model \cite{he2019exact, dooley2013collapse} for an ensemble of two-level systems coupled to a common cavity mode, provided the cavity is occupied by at most one photon. Evolution of this central-spin system has been studied in the context of dynamical nuclear-spin polarization \cite{danon2008tuning,danon2011nuclear,gullans2010dynamic, gullans2013preparation, neder2014theory}, nuclear-spin cooling and coherent electron-nuclear-spin coupling \cite{gangloff2019quantum}, and in transport settings where the nuclear spins are polarized using a spin-polarized current \cite{chesi2015theory,yang2018electrically}.
Another intriguing application of central-spin dynamics is to quantum annealing \cite{kadowaki1998quantum}. 
Quantum annealing correction protocols \cite{pudenz2014error} make direct use of a version of the central-spin system;\footnote{Recent developments \cite{ozfidan2019demonstration} suggest that it will be possible to introduce additional coupling between flux qubits and thus implement our Hamiltonian [Eq.\eqref{eq:Hamiltonian}], where the `penalty qubit' from quantum annealing correction would play the role of the central spin. Given the current structure of the D-wave quantum annealer \cite{denchev2016computational}, this system can realize a central-spin system with ancilla spin $I = 3/2$, while the new Pegasus graph \cite{dattani2019pegasus} could allow for an increased $I$ of up to $15/2$.} understanding how to optimally drive a central-spin system to its steady state could improve these protocols.
Equilibration dynamics of the central-spin system are of special interest, since this system is expected to show a collective enhancement effect \cite{eto2004current,schuetz2012superradiance,chesi2015theory}, related to Dicke superradiance \cite{gross1982superradiance}.

In this paper, we study the dissipative dynamics of a central-spin system when the central spin undergoes spin flips, causing it to relax to its local ground state on a time scale set by the dissipation rate. 
Through a combination of incoherent central-spin flips and a coherent coupling to the ancilla system, the ancilla spins will polarize, eventually reaching a steady state.
When the dissipation is too strong, frequent central-spin flips can be regarded as a projective measurement performed on the central spin, inhibiting unitary dynamics of the coupled system, and preventing ancilla polarization (quantum Zeno effect). 
We will consider a central-spin/ancilla-spin coupling that preserves the total angular momentum along one direction.
For this choice of coupling, ancilla-spin polarization is possible only if the central spin has a finite dissipation rate, allowing angular momentum to be carried away from the ancilla system as it polarizes.
The competition between the required angular-momentum transfer due to dissipation and the Zeno effect indicates that there is an optimal dissipation rate that minimizes the ancilla-spin polarization time. 
Establishing this optimal dissipation rate is a central focus of this paper.

The rest of this paper is organized as follows: in Sec.~\ref{sec:Hamiltonian} we introduce the central-spin Hamiltonian, discuss the dissipative model used to describe dynamics of the ancilla-spin polarization, and present an exact analytical solution for the dynamics of this model.
 In Sec.~\ref{sec:N=1}, we consider the special case of $I=1/2$, relevant for, e.g., two exchange-coupled electron spins. In Sec.~\ref{sec:N>1}, we generalize the $I=1/2$ result and find an expression for the polarization time for any value of $I$. 
 We find a universal relation between the polarization time and the dissipation rate. 
 In Sec.~\ref{sec:physical_applications}, we apply the model [Eq.~\eqref{eq:master_equation_central}] to the problem of fast initialization for two electron-spin qubits in a silicon double quantum dot. Finally, we draw conclusions in Sec.~\ref{sec:discussion}.  Further technical details are given in Appendix \ref{app:master_equation}.

\section{Model and solution}

\begin{figure}
\centering
\includegraphics[width=0.96\columnwidth]{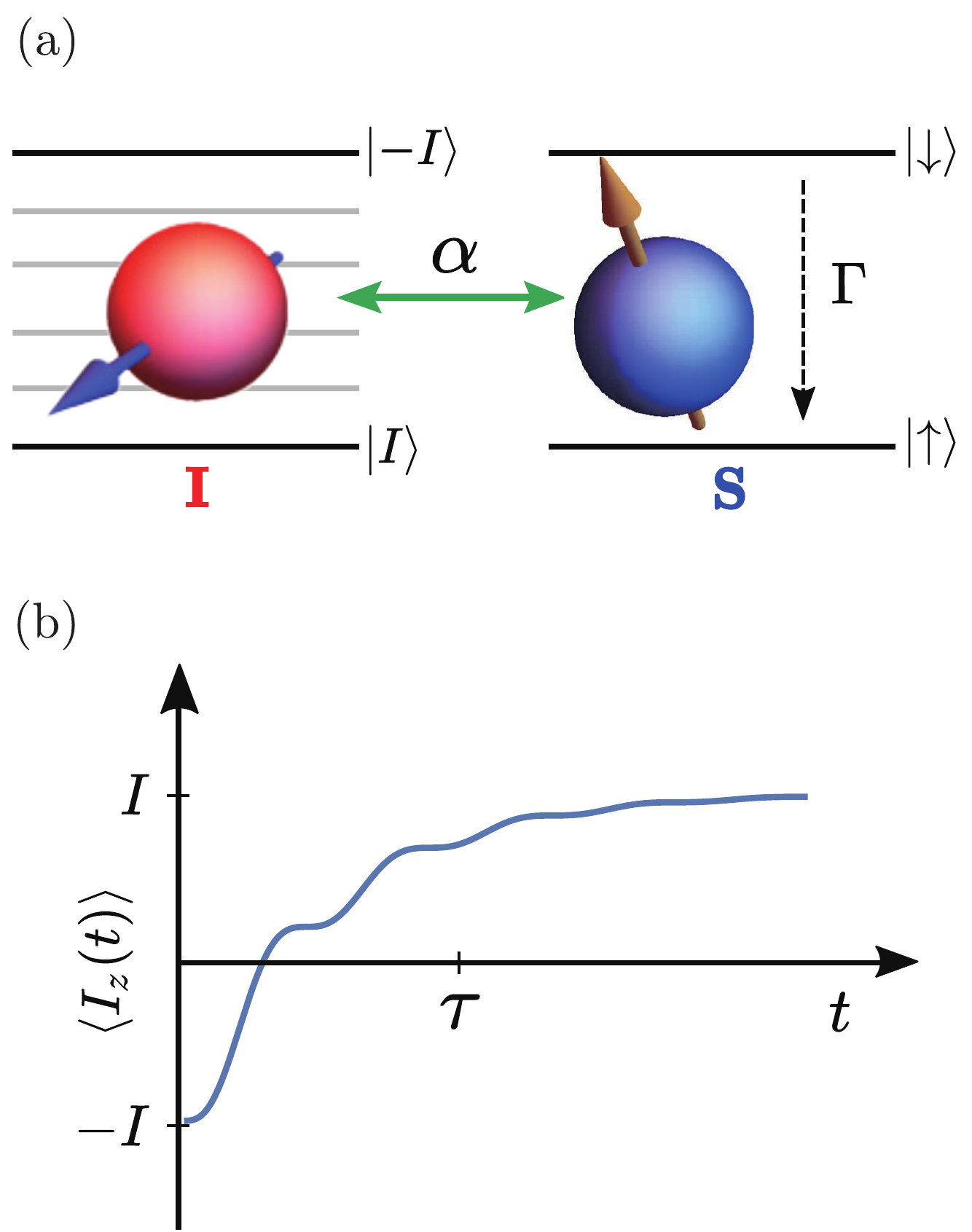}
\caption{(Color online)  (a) The central spin $\mathbf{S}$ couples to the ancilla $\mathbf{I}$ with flip-flop coupling $\alpha$, and undergoes spin-flips ($\downarrow\to\uparrow$) with rate $\Gamma$.  (b) A schematic of the ancilla-spin polarization evolution as a function of time. The ancilla spin reaches the fully-polarized steady-state ($I_z=I$), on average, at the steady-state time $\tau$.}  \label{fig:polarization_sketch}
\end{figure}

A central-spin system can be described in terms of a spin-$1/2$ (the central spin, operator $\mathbf{S}$), and a spin-$I$ (the ancilla spin, operator $\mathbf{I}$), as illustrated in Fig.~\ref{fig:polarization_sketch}(a).  In some realizations of this model, the ancilla spin may describe the collective angular momentum of an ensemble of smaller spins $\mathbf{I}^l$ with uniform coupling: $\mathbf{I}=\sum_l\mathbf{I}^l$ (see Fig.~\ref{fig:css_sketch}).  In other cases, the ancilla may result from a single large spin, giving a $(2I+1)$-dimensional Hilbert space.  Several specific physical realizations of this model are discussed in detail in Section \ref{sec:physical_applications}. The central spin and ancilla spin are coupled with a flip-flop coupling $\alpha$, allowing the coherent exchange of angular momentum. In addition to this coherent coupling, the central spin can also undergo incoherent energy relaxation with a spin-flip rate $\Gamma$ arising, e.g., from coupling to a low-temperature bath [Fig.~\ref{fig:polarization_sketch}(a)].  The angular momentum transferred to the central spin via the incoherent spin flip can then be exchanged via coherent coupling with the ancilla spin. Through this sequence of incoherent spin flips and coherent exchanges, the ancilla spin will eventually reach the fully polarized steady state, $I_z=I$, on a characteristic time scale $\tau$ [the time evolution of $\left< I_z(t)\right>$ is schematically depicted in Fig.~\ref{fig:polarization_sketch}(b)].  The interplay of coherent exchanges (coupling $\alpha$) and incoherent relaxation (rate $\Gamma$) in determining the steady-state time $\tau$ is the main subject of this work.  In particular, when the relaxation rate can be tuned, we would like to establish the optimal choice for $\Gamma$ to minimize the steady-state time $\tau$.   
\label{sec:Hamiltonian}

The state of the combined central spin and ancilla spin is given by the density matrix $\rho$, with evolution governed by the master equation (setting $\hbar=1$):
\begin{equation}\label{eq:master_equation_central}
\dot{\rho}(t)=-i\left[H_\mathrm{CS},\rho(t)\right]+\Gamma\mathcal{D}[S_+]\rho(t). 
\end{equation}
Here, $\mathcal{D}\left[S_+\right]$ is the usual dissipator: 
\begin{equation}
\mathcal{D}[S_+]\rho=S_+\rho S_- -\frac{1}{2}\left\{S_- S_+, \rho \right\},
\end{equation}
where $S_{\pm}=S_x\pm iS_y$.  The Hamiltonian is taken to be
\begin{equation}\label{eq:Hamiltonian}
H_\mathrm{CS}=-\Delta_S S_{z}-\Delta_I I_{z}+J\left[S_{z}I_{z}+\alpha(S_{x}I_{x}+S_{y}I_{y})\right],
\end{equation}
where $\Delta_S(\Delta_I)$ gives the splitting of the central (ancilla) spin and $J$ sets the overall interaction strength.  We will work in dimensionless units where $J=2$ and in a rotating frame, taking advantage of the fact that $\left[H_\mathrm{CS},J_z\right]=0$ ($J_z=S_z+I_z$).  In this rotating frame, the effective Hamiltonian is ($J=2$): 
\begin{equation}\label{eq:HDimensionless}
H=H_\mathrm{CS}+\Delta_IJ_z=-\Delta S_{z}+2S_{z}I_{z}+\alpha(S_{-}I_{+}+S_{+}I_{-}),
\end{equation}
with 
\begin{equation}
\Delta=\Delta_S-\Delta_I.
\end{equation}
The independent parameters of the model are thus the dissipation rate $\Gamma$, the flip-flop coupling $\alpha$, the detuning $\Delta$, and the number of ancilla spins $N$ (or equivalently, the length of the large spin, $I=N/2$).  Equation \eqref{eq:HDimensionless} generally describes an arbitrary XXZ-like interaction.  Depending on the choice of $\alpha$, this Hamiltonian describes a pure Ising-like coupling ($\alpha=0$), Heisenberg interaction ($\alpha=1$), or an XY-like interaction ($\alpha\to\infty$).  The XY-limit is well behaved provided we simultaneously take $\Gamma\to\infty,\Delta\to\infty$, while maintaining finite ratios, $\alpha/\Gamma=\mathrm{const.},\,\Delta/\Gamma=\mathrm{const.}$.  In what follows, the particular form of dissipator in [Eq.~\eqref{eq:master_equation_central}] will be important.  This form can be justified from a microscopic derivation under a generic, but nevertheless restricted, set of conditions [see Appendix \ref{app:master_equation} and Eqs.~\eqref{eq:MarkovConditionPhys}-\eqref{eq:QuasiSecularConditionPhys}, below]. 

The states $\ket{\sigma m}\equiv \ket{\sigma}\otimes \ket{I, m}$ form a convenient basis, where $\sigma=\uparrow,\downarrow$ gives the state of the central spin (eigenstate of $S_z$) and $m$ is the eigenvalue of $I_z$ labelling the state of the ancilla spin.  An initial (non-stationary) state will evolve through a sequence of incoherent central-spin flips (quantum jumps) and coherent angular-momentum exchanges, eventually reaching the final steady state, $\ket{\uparrow I}$.\footnote{That $\ket{\uparrow I}$ is a steady state follows directly from Eq.~\eqref{eq:master_equation_central}, as the projector $\ket{\uparrow I}\bra{\uparrow I}$ commutes with the Hamiltonian, $\left[H_\mathrm{CS},\ket{\uparrow I}\bra{\uparrow I}\right]=0$, and is unaffected by the dissipator, $\mathcal{D}[S_+] \ket{\uparrow I}\bra{\uparrow I}=0$.} The probability per unit time to reach the state $\ket{\uparrow I}$ at time $t$ can be obtained directly from Eq.~\eqref{eq:master_equation_central}: 
\begin{equation}\label{eq:steady-state-probability}
\dot{p}_{\uparrow I}(t)\equiv \mathrm{Tr}\left\{\ket{\uparrow I}\bra{\uparrow I}\dot{\rho}\right\}=\Gamma p_{\downarrow I}(t),
\end{equation}
where
\begin{equation}
p_{\sigma m}(t)=\bra{\sigma m}\rho(t)\ket{\sigma m}
\end{equation}
gives the probability to be in state $\ket{\sigma m}$ and Eq.~\eqref{eq:steady-state-probability} follows from the master equation, Eq.~\eqref{eq:master_equation_central}.  The average time required to reach the steady state is then
\begin{equation}\label{eq:tau}
\tau =\Gamma\int_0^\infty dt \;t\, p_{\downarrow I}(t).
\end{equation}
The steady-state is reached through a sequence of incoherent quantum jumps, so the total time $\tau$ [Eq.~\eqref{eq:tau}] can be re-expressed in terms of a sum over independent average jump times $\tau_m$:
\begin{eqnarray}
\tau &=& \sum_{m=-I}^{I-1}\tau_m,\label{eq:tau-sum}\\
\tau_m &=& \Gamma \int_0^\infty dt\;t\,P(\downarrow m+1,t|\uparrow m,0).\label{eq:tau-m-conditional}
\end{eqnarray}
Here, we have introduced the conditional probability, $P(A,t|B,0)$, to occupy state $A$ at time $t$, for the initial state $B$ at time $t=0$.  
In Sec.~\ref{sec:Bloch_equations}, we find a closed-form analytical expression for the probability $p_{\downarrow m+1}(t)$ for an arbitrary initial state in the subspace of $J_z=m+1/2$.  
Specializing to the initial condition $p_{\uparrow m}(0)=1$ gives $p_{\downarrow m+1}(t)=P(\downarrow m+1,t|\uparrow m,0)$, allowing for an exact calculation of $\tau_m$ as given in Eq.~\eqref{eq:tau-m-conditional}.
Using Eq.~\eqref{eq:tau-sum} we obtain an analytical expression for  $\tau$ which can be minimized for the optimal choice of dissipation rate $\Gamma$.

For the special case of $N=1$ ancilla ($I=1/2$), we evaluate Eq.~\eqref{eq:tau} directly for an initial state having $J_z=0$ in Sec.~\ref{sec:N=1}.
In Sec.~\ref{sec:N>1}, we address the case of multiple ancillas ($N=2I>1$), where a simple choice is to prepare the system in the state $\ket{\uparrow -I}$. 
We evaluate the sum in Eq.~\eqref{eq:tau-sum} explicitly for any $N\ge 1$ ($I\ge 1/2$), analyzing the scaling for a large number of ancillas (large ancilla spin $I$).

\subsection{Bloch equations}
\label{sec:Bloch_equations}

\begin{figure}
\includegraphics[width=0.6\columnwidth]{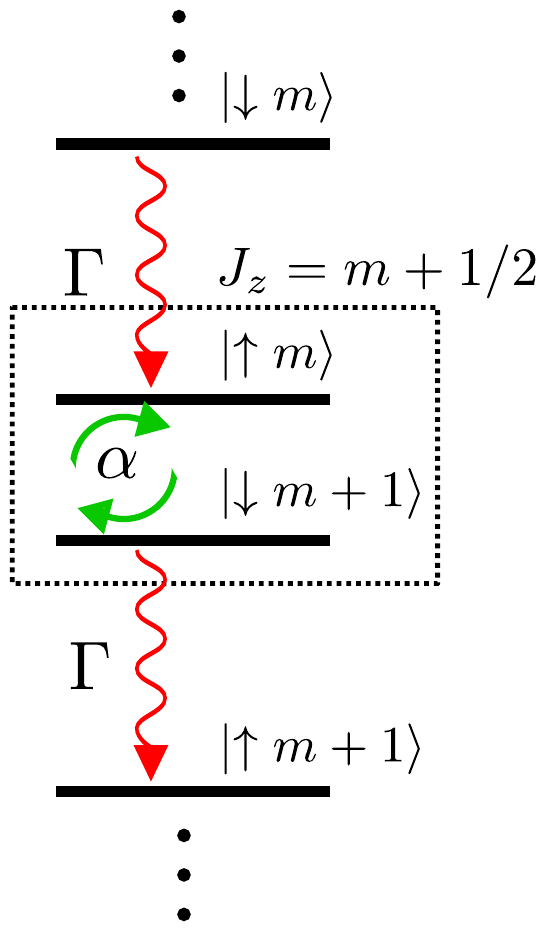}
\caption{(Color online)  The central-spin decay cascade corresponds to quantum jumps between $2\times 2$ blocks, labelled by the total angular momentum $J_z=m+1/2$ (dashed rectangle).}
\label{fig:subspace_dynamics}
\end{figure}

In this section, we find an exact solution to the set of Bloch equations arising from Eq.~\eqref{eq:master_equation_central}.
The equations can be solved exactly provided the initial state lies within a fixed subspace of $J_z=m+1/2$ (see Fig.~\ref{fig:subspace_dynamics}).  Given this initial condition at $t=0$, $\bra{\downarrow m}\rho(t)\ket{\downarrow m}=0$ for all $t>0$ and the Bloch equations for each $J_z=m+1/2$ subspace decouple ($m=-I,-I+1,\ldots,I$).  We then solve the remaining coupled equations for each subspace in terms of four real parameters:
\begin{eqnarray}
p_{\uparrow m}(t) &=& \bra{\uparrow m}\rho(t)\ket{\uparrow m},\\
p_{\downarrow m+1}(t) &=& \bra{\downarrow m+1}\rho(t)\ket{\downarrow m+1},\\
\Omega_{Im}(t) &=& \mathrm{Im}\left\{\bra{\downarrow m+1}\rho(t)\ket{\uparrow m}\right\},\\
\Omega_{Rm}(t) &=& \mathrm{Re}\left\{\bra{\downarrow m+1}\rho(t)\ket{\uparrow m}\right\}.
\end{eqnarray}
The Bloch equations are
\begin{align}
\dot{\mathbf{v}}(t)&=\mathsf{\Lambda}\cdot\mathbf{v}(t);\quad\mathbf{v}=(p_{\uparrow m},p_{\downarrow m+1},\Omega_{Im}, \Omega_{Rm})^T,\label{eq:BlochEquation}\\
\mathsf{\Lambda}&=\begin{pmatrix}
0&0&2\alpha_m&0\\
0&-\Gamma&-2\alpha_m&0\\
-\alpha_m&\alpha_m&-\Gamma/2&-\delta_m\\
0&0&\delta_m&-\Gamma/2
\end{pmatrix}, \label{eq:matrix_diff}
\end{align}
where we have introduced
\begin{eqnarray}
\alpha_m & = & \alpha \sqrt{I(I+1)-m(m+1)},\label{eq:alpha_m_definition}\\
\delta_m & = & \Delta-(2m+1).\label{eq:delta_m_definition}
\end{eqnarray}

The secular equation, $\mathrm{det}\left(\mathsf{\Lambda}-\lambda_j\mathsf{1}\right)=0$, is a quartic equation for the eigenvalues $\lambda_j$, but this can be rewritten as a quadratic equation in terms of $x=\left(\lambda_j+\Gamma/2\right)^2$, which can be solved directly giving the four eigenvalues of $\mathsf{\Lambda}$:
\begin{eqnarray}
\lambda_{1,2}&=&-\frac{1}{2}\left(\Gamma\pm\sqrt{2}\sqrt{\sqrt{\Xi_m^4+\Gamma^2\delta_m^2}+\Xi_m^2}\right),\label{eq:lambda12}\\
\lambda_{3,4}&=&-\frac{1}{2}\left(\Gamma\pm i\sqrt{2}\sqrt{\sqrt{\Xi_m^4+\Gamma^2\delta_m^2}-\Xi_m^2}\right),\label{eq:lambda34}
\end{eqnarray}
with
\begin{eqnarray}
\Xi_m^2=\left(\Gamma/2\right)^2-4 \alpha_m^2-\delta_m^2.
\end{eqnarray}
The eigenvalues $\lambda_{1,2}$ are always purely real, leading to pure exponential decay for any $\Gamma\ne 0$.  For $\Gamma\delta_m\ne 0$, $\lambda_{3,4}$ will always have a finite imaginary component, leading to oscillatory behavior.  A special case occurs for $\delta_m=0$, where $\lambda_{3,4}$ become purely real for $\Gamma>4\alpha_m$.  This case is realized, e.g., for $I=1/2$, $m=-1/2$, $\Delta=0$ [see Eq.~\eqref{eq:delta_m_definition}].  This particular case is discussed in Sec.~\ref{sec:N=1}.

The right (left) eigenvectors $\mathbf{v}_{R(L)j}$ can be found from
\begin{eqnarray}
\mathsf{\Lambda}\cdot\mathbf{v}_{Rj} & = & \lambda_j\mathbf{v}_{Rj},\\
\mathbf{v}_{Lj}^T\cdot\mathsf{\Lambda} & = & \lambda_j\mathbf{v}_{Lj}^T,\\
\mathbf{v}_{L i}^T\cdot\mathbf{v}_{R j} & = &\delta_{ij};\quad j=1,2,3,4.\label{eq:EigenvectorsNormalization}
\end{eqnarray}
With the normalization given in Eq.~\eqref{eq:EigenvectorsNormalization}, we can define projection matrices $\mathsf{\Pi}_j$ in terms of an outer product:\begin{equation}
\mathsf{\Pi}_j = \mathbf{v}_{Rj}\cdot\mathbf{v}_{Lj}^T.
\end{equation}

The formal solution to Eq.~\eqref{eq:BlochEquation} is then
\begin{equation}
\mathbf{v}(t) = \sum_j e^{\lambda_j t}\mathsf{\Pi}_j\cdot\mathbf{v}(0),
\end{equation}
and we can find, e.g., $p_{\downarrow m+1}$, from
\begin{equation}\label{eq:p_dn_general}
p_{\downarrow m+1}(t) = \sum_j e^{\lambda_j t}\mathbf{p}_{\downarrow m+1}^T\cdot\mathsf{\Pi}_j\cdot\mathbf{v}(0), 
\end{equation}
with
\begin{equation}
\mathbf{p}_{\downarrow m+1}=\left(0,1,0,0\right)^T.
\end{equation}
For a given initial condition, $\mathbf{v}(0)$, we can use Eq.~\eqref{eq:p_dn_general} to evaluate the average time to reach the state $\ket{\uparrow m+1}$:
\begin{equation}\label{eq:tau_s_definition}
\tau_m = \int_0^\infty dt \Gamma t p_{\downarrow m+1}(t).
\end{equation}
When $\Gamma\ne 0$, we have $\mathrm{Re}\left[\lambda_j\right]<0$ for all eigenvalues [see Eqs.\,\eqref{eq:lambda12} and \eqref{eq:lambda34}].  In this case, the integral in Eq.~\eqref{eq:tau_s_definition} is well defined and we insert Eq.~\eqref{eq:p_dn_general} into Eq.~\eqref{eq:tau_s_definition} to find
\begin{equation}\label{eq:tau_s_evaluated}
\tau_m = \sum_j \frac{1}{\lambda_j^2}\mathbf{p}_{\downarrow m+1}^T\cdot\Pi_j\cdot\mathbf{v}(0).
\end{equation}

A general initial state, restricted to the subspace of $J_z=m+1/2$ (dashed box in Fig.~\ref{fig:subspace_dynamics}) is given by
\begin{equation}
\mathbf{v}(0) = \left(p_{\uparrow m}(0),1-p_{\uparrow m}(0),\Omega_{Im}(0),\Omega_{Rm}(0)\right)^T.
\end{equation}
Inserting this initial state into Eq.~\eqref{eq:tau_s_evaluated} gives the average time required to reach the state $\ket{\uparrow m+1}$:
\begin{equation}\label{eq:tau_m}
\tau_m=\frac{2}{\Gamma}+\frac{\Gamma^2+4\delta_m^2}{4\alpha_m^2\Gamma}p_{\uparrow m}(0)+\frac{\Omega_{Im}(0)}{\alpha_m}-\frac{2\delta_m}{\alpha_m\Gamma}\Omega_{Rm}(0).
\end{equation}

\section{Single ancilla spin ($N=1;\,I=1/2$)}
\label{sec:N=1}
\begin{figure}
\centering
\includegraphics[width=0.96\columnwidth]{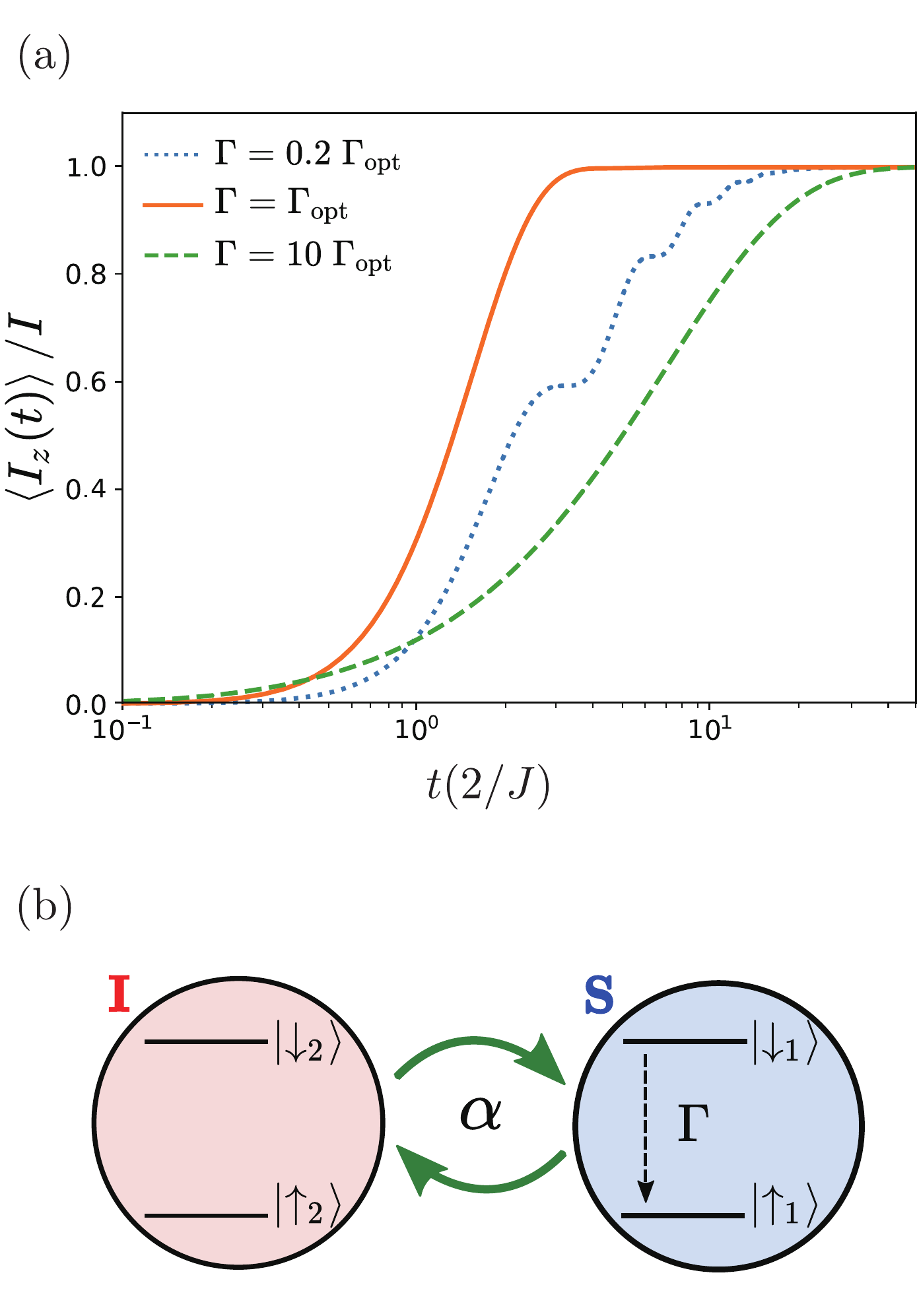}
\caption{(Color online)  (a) Ancilla-spin polarization $\langle I_z(t) \rangle $ for the single-ancilla-spin model ($N=1, I=1/2$). Here, we have taken $\Delta=\Delta_\mathrm{S}-\Delta_\mathrm{I}=0$ and $\alpha=1$.  For these parameters, the three values of $\Gamma$ indicated correspond to the underdamped regime ($\Gamma<\Gamma_{\mathrm{opt}}=4\alpha$, blue dotted line), the optimal value ($\Gamma=\Gamma_{\mathrm{opt}}=4\alpha$, orange solid line), and the overdamped (quantum Zeno) regime ($\Gamma>\Gamma_{\mathrm{opt}}=4\alpha$, green dashed line).   (b) A double quantum dot. The central spin $\mathbf{S}$ (right dot) decays to its ground state with rate $\Gamma$ while exchanging angular momentum with the ancilla spin $\mathbf{I}$ (left dot) with a flip-flop coupling $\alpha$.}  \label{fig:single_spin}
\end{figure}

The simplest case of a single ancilla spin-1/2 $(N=1,I=1/2)$ is directly relevant to, e.g., two electron spins in a double quantum dot [Fig.~\ref{fig:single_spin}(b)] coupled by a Heisenberg exchange coupling $J$ ($\alpha=1$), or to an electron spin bound to a phosphorus donor impurity in silicon, coupled to the $^{31}$P nuclear spin via the Fermi contact interaction $A$ ($\alpha=1,\,J\to A$).  Both of these systems show promise as elements for quantum information processing \cite{morello2010single, watson2018programmable, zajac2018resonantly, he2019two}, where it is important to have a fast and high-fidelity reset to the fiducial initial state $\ket{\uparrow_1\uparrow_2}$ for both the central-spin ($\sigma_1$) and the ancilla ($\sigma_2$), even when only the central spin undergoes direct spin relaxation [Fig.~\ref{fig:single_spin}(b)].  By optimizing the dissipation rate $\Gamma$, the time scale for this reset (the steady-state time $\tau$) can be minimized, improving the rate at which fresh ancillas can be prepared.

Before analyzing the steady-state time, we consider the ancilla-spin dynamics:
\begin{equation}\label{eq:Izt}
\left<I_z(t)\right>=\sum_{\sigma m}m p_{\sigma m}(t).
\end{equation}  
In Sec.~\ref{sec:Bloch_equations}, we have shown that $p_{\sigma m}(t)$ can be found exactly analytically for this model, whenever the initial state lies in a subspace of fixed $J_z$.  Here, we consider the singlet initial state, which can be prepared efficiently for two electron spins in a double quantum dot, taking advantage of Pauli exclusion \cite{petta2005coherent}:
\begin{equation}
\rho (0) = \ket{S}\bra{S}; \qquad  \ket{S}=\frac{1}{\sqrt{2}}\left(\left |\downarrow_1\uparrow_2\right>-\left|\uparrow_1\downarrow_2\right>\right).
\label{eq:initial_singlet}
\end{equation}
The resulting dynamics are shown for $\Delta=0,\,\alpha=1$, and three values of $\Gamma$ in Fig.~\ref{fig:single_spin}(a).  There are two striking features: (i) the dynamics are qualitatively different for weak dissipation (showing oscillations) and for strong dissipation (showing no oscillations), and (ii) the steady-state time (time to reach $\left<I_z(t)\right>/I=1$) is a non-monotonic function of the dissipation rate $\Gamma$.  In particular, there is an optimal choice for the dissipation rate, $\Gamma=\Gamma_\mathrm{opt}$, that minimizes the steady-state time [orange curve in Fig.~\ref{fig:single_spin}(a)].

To better understand the dynamics, it is useful to consider one of the terms entering Eq.~\eqref{eq:Izt}.  Following the procedure described in Sec.~\ref{sec:Bloch_equations}, and for the singlet initial state, we find (for $\Delta=0$):
\begin{align}
p_{\downarrow I}(t) &=  \frac{e^{-\frac{\Gamma}{2} t}}{8\Xi^2} \left[\Gamma^2  \cosh\left(\Xi t\right) -2\Gamma \Xi \sinh\left(\Xi t\right)-16\alpha^2\right],\label{eq:eom}\\
\Xi&=\frac{1}{2}\sqrt{\Gamma ^2-16\alpha^2}\label{eq:d}.
\end{align}
The qualitative behavior of this function changes from oscilliatory when $\Gamma<4\alpha$ [$\mathrm{Im}\left(\Xi\right)\ne 0$] to purely damped when $\Gamma>4\alpha$ [$\mathrm{Im}\left(\Xi\right)= 0$]. We can therefore distinguish three different regimes in analogy with the damped harmonic oscillator: (i) the underdamped regime ($\Gamma<4\alpha$), where the ancilla spin undergoes many oscillations before reaching the steady state, (ii) the overdamped regime ($\Gamma>4\alpha$), where the dissipation drives the system into the quantum Zeno regime and where the steady state is reached without any oscillations, and (iii) the critically damped regime ($\Gamma=4\alpha$).

For an arbitrary initial state in the subspace $J_z=0$, the steady-state time is given by Eq.~\eqref{eq:tau_m} when we set $m=-1/2,\,I=1/2$, resulting in:
\begin{equation}\label{eq:tauIhalf}
\tau=\frac{2}{\Gamma}+\frac{\left(\Gamma^2+4\Delta^2\right)P_{\uparrow\downarrow}}{4\alpha^2\Gamma}+\frac{\mathrm{Im}\left[(\Gamma-i 2\Delta)\Omega\right]}{\alpha\Gamma},
\end{equation}
where
\begin{eqnarray}
P_{\uparrow\downarrow}&=&\bra{\uparrow_1\downarrow_2}\rho(0)\ket{\uparrow_1\downarrow_2},\\
\Omega &=& \bra{\downarrow_1\uparrow_2}\rho(0)\ket{\uparrow_1\downarrow_2},
\end{eqnarray}
and for a physical (positive) state, $\mathrm{det}\left[\rho(0)\right]\ge 0$ implies
\begin{equation}
\left|\Omega\right|\le\sqrt{P_{\uparrow\downarrow}(1-P_{\uparrow\downarrow})}. 
\end{equation}
From Eq.~\eqref{eq:tauIhalf}, we have $\tau\to\infty$ as $\Gamma\to\infty$ for any initial state with $P_{\uparrow\downarrow}\ne 0$ and for any finite $\alpha$, since the central spin will be continuously projected onto its local ground state $\ket{\uparrow}$, inhibiting coherent exchange with the ancilla (quantum Zeno effect \cite{misra1977zeno,peres1980zeno}).  In the opposite limit ($\Gamma\to 0$), the steady-state time still diverges, $\tau\to \infty$, since angular momentum cannot be carried away without dissipation in this model.  There will be an optimal intermediate value of $\Gamma$ that minimizes $\tau$.  The resulting minimal (optimal) steady-state time $\tau_\mathrm{min}$ and optimal choice for the rate $\Gamma_\mathrm{opt}$ are:
\begin{eqnarray}
\tau_\mathrm{min} & = & 2\frac{\left(2\alpha^2+\Delta^2 P_{\uparrow\downarrow}\right)}{\alpha^2\Gamma_\mathrm{opt}}+\frac{\mathrm{Im}\left[\left(\Gamma_\mathrm{opt}-i2\Delta\right)\Omega\right]}{\alpha\Gamma_\mathrm{opt}},\label{eq:tau-opt-Ihalf}\\
\Gamma_\mathrm{opt} & = & 2\left\{\frac{2\alpha\left[\alpha-\Delta\mathrm{Re}\left(\Omega\right)\right]+\Delta^2 P_{\uparrow\downarrow}}{P_{\uparrow\downarrow}}\right\}^{1/2}.\label{eq:Gamma-opt-Ihalf}
\end{eqnarray}

For specific pure-state initial conditions $\rho(0)=\ket{\psi(0)}\bra{\psi(0)}$, Eqs.~\eqref{eq:tau-opt-Ihalf} and \eqref{eq:Gamma-opt-Ihalf} give simple results. For, e.g., $\ket{\psi(0)}=\ket{\uparrow_1\downarrow_2}$, we find
\begin{equation}
\tau_\mathrm{min}=\frac{\sqrt{\Delta^2+2\alpha^2}}{\alpha^2};\quad\Gamma_\mathrm{opt}=2\sqrt{\Delta^2+2\alpha^2}.\label{eq:GammaOpt2Spins}
\end{equation}
Consistent with the analogy to a damped oscillator, the minimal steady-state time is found when the dissipation rate is resonantly synchronized with the coherent dynamics, $\Gamma_\mathrm{opt}\propto\mathrm{max}\left(\alpha,|\Delta|\right)$. 

Alternatively, for $\ket{\psi(0)}=\left(\ket{\uparrow_1\downarrow_2}+e^{i\phi}\ket{\downarrow_1\uparrow_2}\right)/\sqrt{2}$, we have (for $\Delta=0$):
\begin{equation}\label{eq:tau-opt-Delta-zero}
\tau_\mathrm{min}=\frac{2+\sin\phi}{2\alpha};\quad\Gamma_\mathrm{opt}=4\alpha\quad (\Delta=0).
\end{equation}
For either the singlet ($\phi=\pi$) or triplet ($\phi=0$) initial states, the minimal steady-state time is $\tau_\mathrm{min}=1/\alpha$, but this time is reduced by half ($\tau_\mathrm{min}=1/2\alpha$) for an initial state with $\phi=-\pi/2$.  For electron spin qubits, it is often possible to prepare an arbitrary state in the $J_z=0$ subspace by rapidly preparing a spin singlet via Pauli exclusion, followed by a controlled phase gate \cite{veldhorst2015two, watson2018programmable, zajac2018resonantly}.  
Equation \eqref{eq:tau-opt-Delta-zero} shows that the local ground state $\ket{\uparrow_1\uparrow_2}$ (the fiducial initial state for a computation) can be reached twice as fast by first preparing the `correct' initial state when only one spin undergoes direct spin relaxation.

\section{Multiple ancilla spins ($N\ge 1$; $I\ge 1/2$)}
\label{sec:N>1}

\begin{figure}
\centering
\includegraphics[width=0.96\columnwidth]{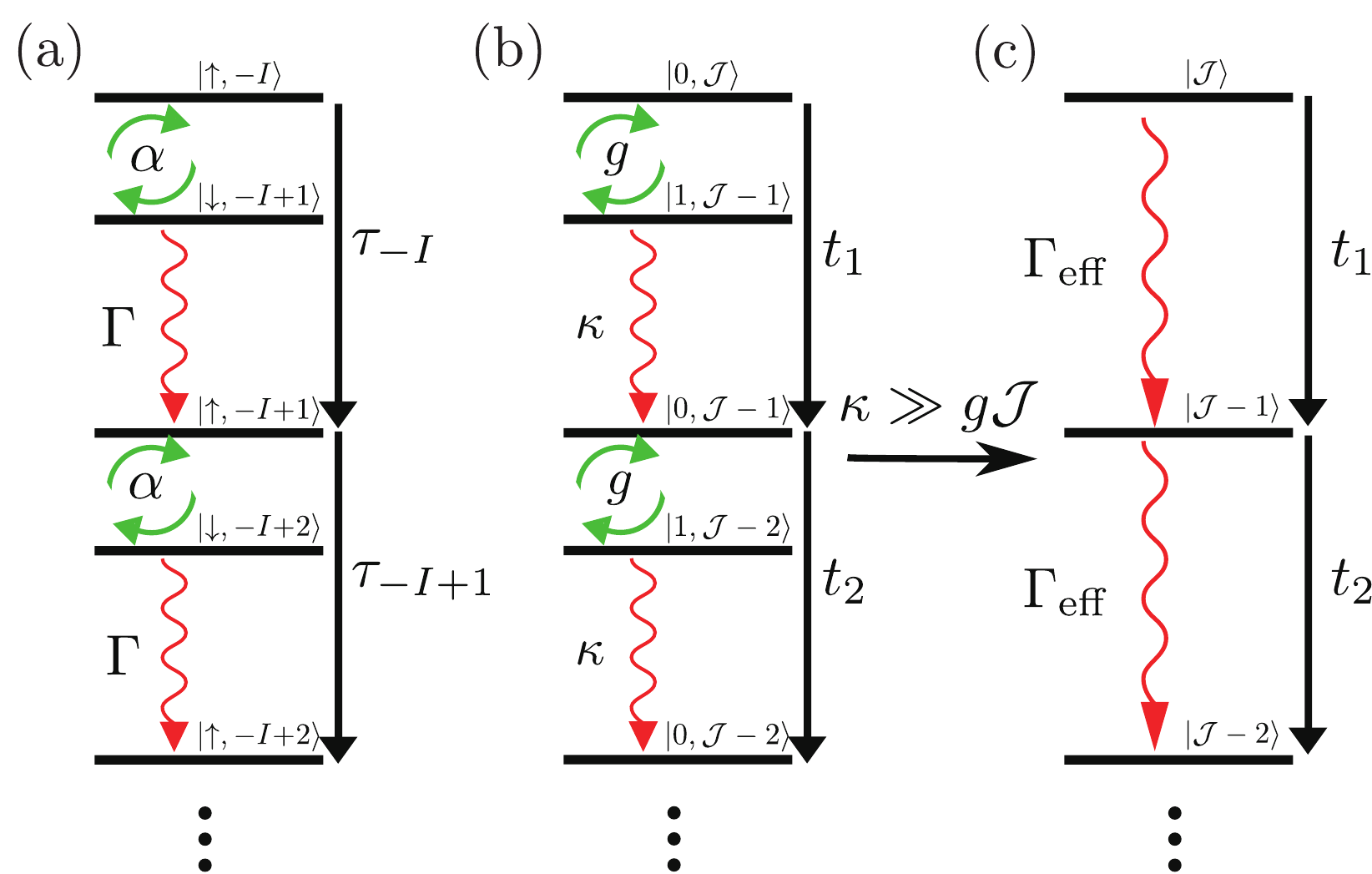}
\caption{(Color online)  (a) Central-spin decay cascade, controlled by the flip-flop coupling $\alpha$ and spin-flip rate $\Gamma$.  The average time to reach the state $\ket{\uparrow,m+1}$ from $\ket{\uparrow,m}$ is $\tau_m$. (b) Decay cascade for an ensemble of $N=2\mathcal{J}$ two-level atoms with cavity coupling $g$ and cavity decay rate $\kappa$.  The state $\ket{n,\mathcal{J}-s}$ describes $n$ photons in the cavity and an atomic Dicke state having collective angular momentum $\mathcal{J}_z=\mathcal{J}-s$ after $s$ photons are emitted.  The time for the $s^\mathrm{th}$ photon to be emitted is $t_s$.  (c) In the bad-cavity limit, $\kappa \gg \mathcal{J}g$, the dynamics are approximated by a sequence of independent incoherent jump events, with an effective rate $\Gamma_\mathrm{eff}\propto g^2/\kappa$ \cite{gross1982superradiance}. }\label{fig:polarization_dynamics}
\end{figure}

The case of $N=2I\ge 1$ ancilla spins results in the decay cascade illustrated in Fig.~\ref{fig:polarization_dynamics}(a).  As described above, the decay proceeds through a sequence of coherent angular-momentum exchanges (due to the flip-flop coupling $\alpha$) and incoherent central-spin flips (with rate $\Gamma$).  There is a direct correspondence between this (central-spin) model and the well-studied model of Dicke superradiance leading to a decay cascade for a collection of $N$ two-level atoms coupled to a cavity via a coherent coupling $g$ and with a cavity decay rate $\kappa$ [Fig.~\ref{fig:polarization_dynamics}(b)].  
The correspondence becomes exact if the cavity only ever contains $n=0,1$ photons, in which case we make the identification $\ket{0/1,\mathcal{J}-s}\leftrightarrow\ket{\uparrow/\downarrow,-I+s}$, with $s=0,1,2\ldots 2I\;(2\mathcal{J})$.  Here, $\ket{n,\mathcal{J}-s}$ describes a state with $n$ cavity photons and a collective (Dicke) state of the atomic ensemble having collective angular momentum $\mathcal{J}_z=\mathcal{J}-s$.\footnote{Dynamics under the usual Tavis-Cummings Hamiltonian [$H_\mathrm{TC}=g\left(\mathcal{J}_+a+\mathcal{J}_-a^\dagger\right)$] and cavity decay at rate $\kappa$ (projected onto the subspace of $n=a^\dagger a=0,1$ cavity photons) can be realized from the central-spin equation of motion [Eq.~\eqref{eq:master_equation_central}] in the limit $\Gamma\to\infty,|\Delta|\to\infty,\alpha/\Gamma\to g/\kappa=\mathrm{const.}$} In the bad-cavity limit ($\kappa\gg g\mathcal{J}$), photons leak out of the cavity as quickly as they are produced through the decaying atomic ensemble, so the state with $n=2$ photons is never reached.  In this limit, the photon emission dynamics can be found in terms of an equivalent decay cascade that incorporates the coupling $g$ and decay rate $\kappa$ into a single rate $\Gamma_\mathrm{eff}\propto g^2/\kappa$ [Fig.~\ref{fig:polarization_dynamics}(c)].  The dynamics of photon emission for the model shown in Fig.~\ref{fig:polarization_dynamics}(c) have been analyzed in detail, showing interesting collective-enhancement effects that lead to an accelerated approach to the steady state due to superradiance (see, e.g., Ref.~\onlinecite{gross1982superradiance} for a review).  Analogous collective-enhancement effects are inherited by the central-spin model \cite{eto2004current, kessler2012dissipative, chesi2015theory, he2019exact, fang2020superradiant}, but the scaling with large $N=2I$ can be fundamentally different.  As we will show below, the bad-cavity limit is analogous to the quantum Zeno regime for the central-spin model ($\Gamma\gg I\alpha$), where the steady-state time increases as the associated dissipation rate ($\kappa$ or $\Gamma$) is increased.  Working in this limit allows for a convenient controlled description of superradiance, but it misses the potential to further accelerate the decay cascade by optimizing $\Gamma$.  

The average time $\tau_m$ for the transition $\ket{\uparrow m}\to \ket{\uparrow m+1}$ can be found directly by integrating the Bloch equations resulting from Eq.~\eqref{eq:master_equation_central}.  This analysis is carried out explicitly in Sec.~\ref{sec:Bloch_equations}. Inserting the initial condition $p_{\uparrow m}(0)=1$ into Eq.~\eqref{eq:tau_m}, we find
\begin{equation}\label{eq:tau_m_init}
\tau_m = \frac{2}{\Gamma}+\frac{\Gamma^2+4\delta_m^2}{4\alpha_m^2\Gamma}.
\end{equation}
The parameters $\alpha_m,\delta_m$ are given in Eqs.~\eqref{eq:alpha_m_definition}-\eqref{eq:delta_m_definition}, but we rewrite them here for convenience:
\begin{eqnarray}
\alpha_m & = & \alpha\sqrt{I(I+1)-m(m+1)},\\
\delta_m & = & \Delta-\left(2m+1\right).
\end{eqnarray} 
We can sum the individual jump times, given by Eq.~\eqref{eq:tau_m_init}, to find the total steady-state time $\tau$ [as in Eq.~\eqref{eq:tau-sum}].  The result is given by an exact analytic expression:
\begin{multline}
\tau=2\frac{\left[(\Gamma/2)^2+(2I+1)^2+\Delta^2\right]}{\alpha^2\Gamma}\frac{h(2I)}{2I+1}+\\
+\frac{4I(\alpha^2-2)}{\alpha^2\Gamma},\label{eq:tau_tilde}
\end{multline}
where $h(2I)$ is the harmonic series:
\begin{equation}
h(N_0)=\sum\limits_{n=1}^{N_0} \frac{1}{n}.
\end{equation}
In the large-$I$ limit, the harmonic series has the asymptotic expression
\begin{equation}
h(2I)\sim \ln 2I+\gamma;\quad\left(I\rightarrow\infty\right),
\end{equation} where $\gamma\approx 0.57$ is the Euler-Mascheroni constant.

\subsection{Scaling with $N=2I$}

We can recover the result applicable to the decay cascade for an atomic ensemble [shown in Fig.~\ref{fig:polarization_dynamics}(b)] by first taking the limits $\Gamma\to \infty$, $\alpha\to \infty$, $|\Delta|\to \infty$, $\alpha/\Gamma=\mathrm{const.},\alpha/\Delta=\mathrm{const.}$ [appropriate for an XY-like interaction in Eq.~\eqref{eq:HDimensionless}], then setting $\alpha/\Gamma\to g/\kappa$, $\alpha/\Delta\to g/\Delta$.  In the limit $2I=2\mathcal{J}=N\to\infty$, this gives:
\begin{equation}\label{eq:DickelnNdivN}
\tau \sim \frac{\left(\kappa^2+\Delta^2\right)}{2g^2\kappa}\frac{\ln N}{N};\quad N\to \infty.
\end{equation}  
For $\Delta=0$, this result matches, e.g., with the analysis of Ref.~\onlinecite{gross1982superradiance} for the model of Dicke superradiance. A striking feature of this model is that the steady-state time actually \emph{decreases} with increasing $N$ (a consequence of the collective enhancement).  For any nonzero detuning $\Delta\ne 0$, the steady-state time $\tau$ can be minimized by choosing $\kappa=\Delta$.  In most contexts, however, it is likely to be simpler to arrange for $\Delta=0$ to minimize $\tau$ for any fixed $\kappa$.  

The situation is quite different for a central-spin system with a finite Ising-like contribution $\sim S_zI_z$.  This contribution is unavoidable for, e.g., spins coupled via a Heisenberg-like interaction through exchange or a Fermi contact hyperfine coupling.  In this case, the large-$I$ limit of Eq.~\eqref{eq:tau_tilde} gives
\begin{equation}\label{eq:tau-asympI}
\tau \sim \frac{4I\ln(2I)}{\alpha^2\Gamma}+\mathcal{O}(I);\quad I\to \infty.
\end{equation}
In stark contrast with the decay cascade for an atomic ensemble (or central-spin system with an XY-coupling), the more general central-spin system will show a steady-state time that grows super-linearly with $2I=N$ (assuming that the coupling parameters $J$ and $ \alpha$ are independent of $N$).  This is even worse than what one would expect for uncorrelated incoherent spin-flips with each of the $N$ ancilla spins.  The situation improves, however, if we treat the spin-flip rate $\Gamma$ as a parameter that can be optimized for each fixed $I$.

Minimizing Eq.~\eqref{eq:tau_tilde} with respect to variations in $\Gamma$ gives:
\begin{eqnarray}
\tau_\mathrm{min} &=&\frac{h(2I)}{\alpha^2(2I+1)}\Gamma_\mathrm{opt}(\alpha, 
\Delta, I),\label{eq:tau_tilde_opt}\\
\Gamma_\mathrm{opt} &=& 2\left[\frac{2I(2I+1)(\alpha^2-2)}{h(2I)}+(2I+1)^2+\Delta^2\right]^{1/2}.\label{eq:Gamma_opt}
\end{eqnarray}

In the limit of large $N=2I$, Eqs.~\eqref{eq:tau_tilde_opt} and \eqref{eq:Gamma_opt} then give:
\begin{equation}\label{eq:tau_opt_asymp}
\tau_\mathrm{min}\sim \frac{2}{\alpha^2}\ln(2I)+\frac{2\gamma}{\alpha^2};\,\,\Gamma_\mathrm{opt}\sim 2(2I+1);\,\,(I\to \infty).
\end{equation}
The optimized steady-state time grows only very slowly (logarithmically) with $N=2I$.  This optimization is only practical as long as the spin-flip rate can be increased in proportion with $I$ [restoring to dimensionful units, $\Gamma_\mathrm{opt}\sim (2I+1)/J$].  The scaling improves substantially if the detuning $\Delta$ is large:
\begin{eqnarray}
\tau_\mathrm{min}\sim \frac{2|\Delta|}{\alpha^2}\frac{\ln(2I)}{2I};\quad\Gamma_\mathrm{opt}\sim 2|\Delta|,\label{eq:tau_opt_Delta}\label{eq:opt_large_detuning}\\
\left[|\Delta|\gg 2I\gg \mathrm{max}\left(1,e^{|\alpha^2-2|}\right)\right].
\end{eqnarray}     
Thus, the central-spin system will show a decreasing steady-state time with increasing $N=2I$ over some intermediate regime.

\begin{figure}
\centering
\includegraphics[width=\columnwidth]{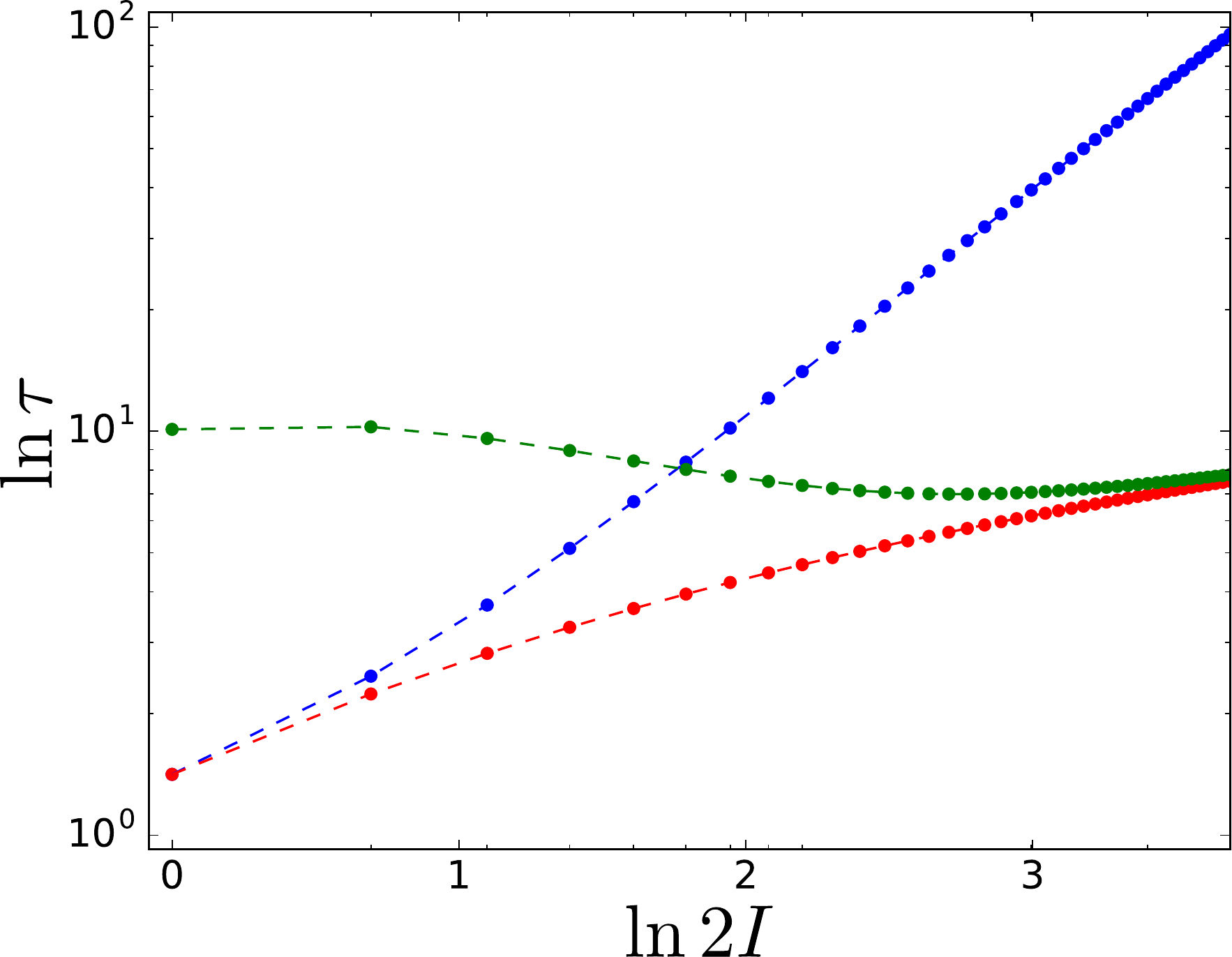}
\caption{(Color online)  The steady-state time $\tau$ [Eq.~\eqref{eq:tau_tilde}] for $\alpha=1$, $\Delta=0$, with fixed spin-flip rate $\Gamma=\Gamma_\mathrm{opt}(\alpha, \Delta, I=1/2)$ (blue dots) and the optimized steady-state time $\tau=\tau_\mathrm{min}$ [Eqs.~\eqref{eq:tau_tilde_opt},\eqref{eq:Gamma_opt}] for $\alpha=1,\,\Delta=0$ (red dots) and for $\alpha=1,\,\Delta=10$ (green dots).  Each dot corresponds to a different value of $I$ in half-integer steps ($I=1/2,1,3/2,\ldots$) and the dashed lines are a linear interpolation between adjacent dots.}\label{fig:diff_scaling}
\end{figure}
To summarize, the interesting scaling limits for $\tau$ (and $\tau_\mathrm{min}$) are illustrated in Fig.~\ref{fig:diff_scaling}.  For $\Delta=0,\,\alpha=1$, the unoptimized steady-state time $\tau$ grows super-linearly with increasing $N=2I$ [blue points, Eq.~\eqref{eq:tau_tilde}, with asymptotics described by Eq.~\eqref{eq:tau-asympI}].  In contrast, choosing $\Gamma=\Gamma_\mathrm{opt}$ for each $I$ leads to a slow logarithmic increase in $\tau=\tau_\mathrm{min}$ [red points, Eq.~\eqref{eq:tau_tilde_opt}, with asymptotics given in Eq.~\eqref{eq:tau_opt_asymp}].  For $\Delta=10,\,\alpha=1$, there is an intermediate regime where $\tau_\mathrm{min}$ initially decreases with increasing $I$ [green points, Eq.~\eqref{eq:tau_tilde_opt}, with the decreasing portion described by Eq.~\eqref{eq:tau_opt_Delta}].   

\subsection{Universal relation}

\begin{figure}
\centering
\includegraphics[width=0.96\columnwidth]{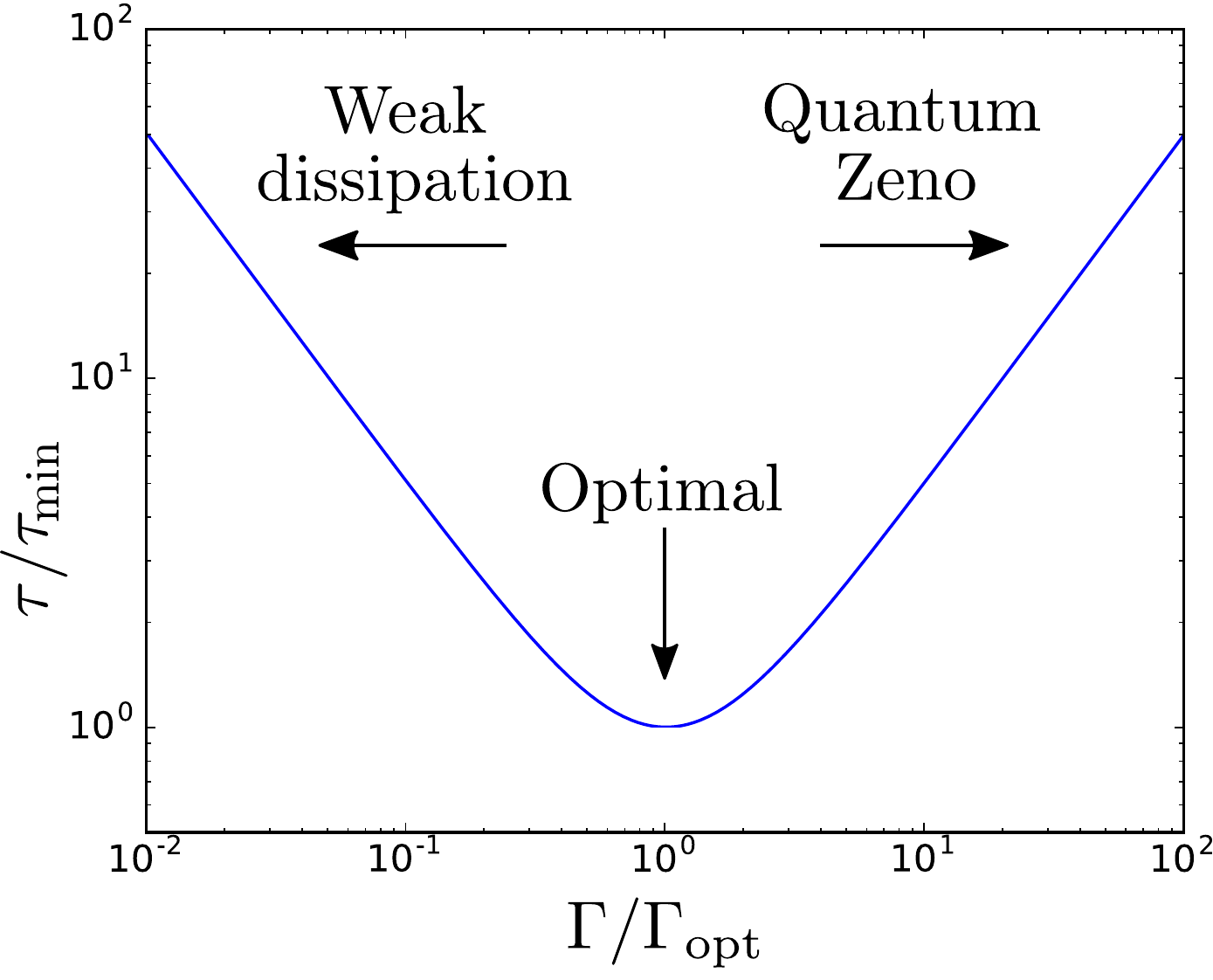}
\caption{(Color online)  Universal relation for $\tau$ vs.~$\Gamma$ [Eq~\eqref{eq:universal}]. For $\Gamma\ll \Gamma_\mathrm{opt}$, the system is in the weak-dissipation regime, while for $\Gamma\gg \Gamma_\mathrm{opt}$ the system is in the quantum Zeno regime. The minimal steady-state time $\tau=\tau_\mathrm{min}$ is found for $\Gamma=\Gamma_\mathrm{opt}$.}\label{fig:universal_curve}
\end{figure}

It follows directly from the form of $\tau$ given in Eq.~\eqref{eq:tau_tilde} that we can write the $\Gamma$-dependence as a universal curve in terms of the scaled parameters $\tau/\tau_\mathrm{min}$ and $\Gamma/\Gamma_\mathrm{opt}$ (see Fig.~\ref{fig:universal_curve}):
\begin{equation}
\frac{\tau}{\tau_\mathrm{min}}=\frac{1}{2}\left(\frac{\Gamma}{\Gamma_\mathrm{opt}}+\frac{\Gamma_\mathrm{opt}}{\Gamma}\right).\label{eq:universal}
\end{equation}

\section{Physical applications}
\label{sec:physical_applications}
\begin{figure}
\includegraphics[width=0.96\columnwidth]{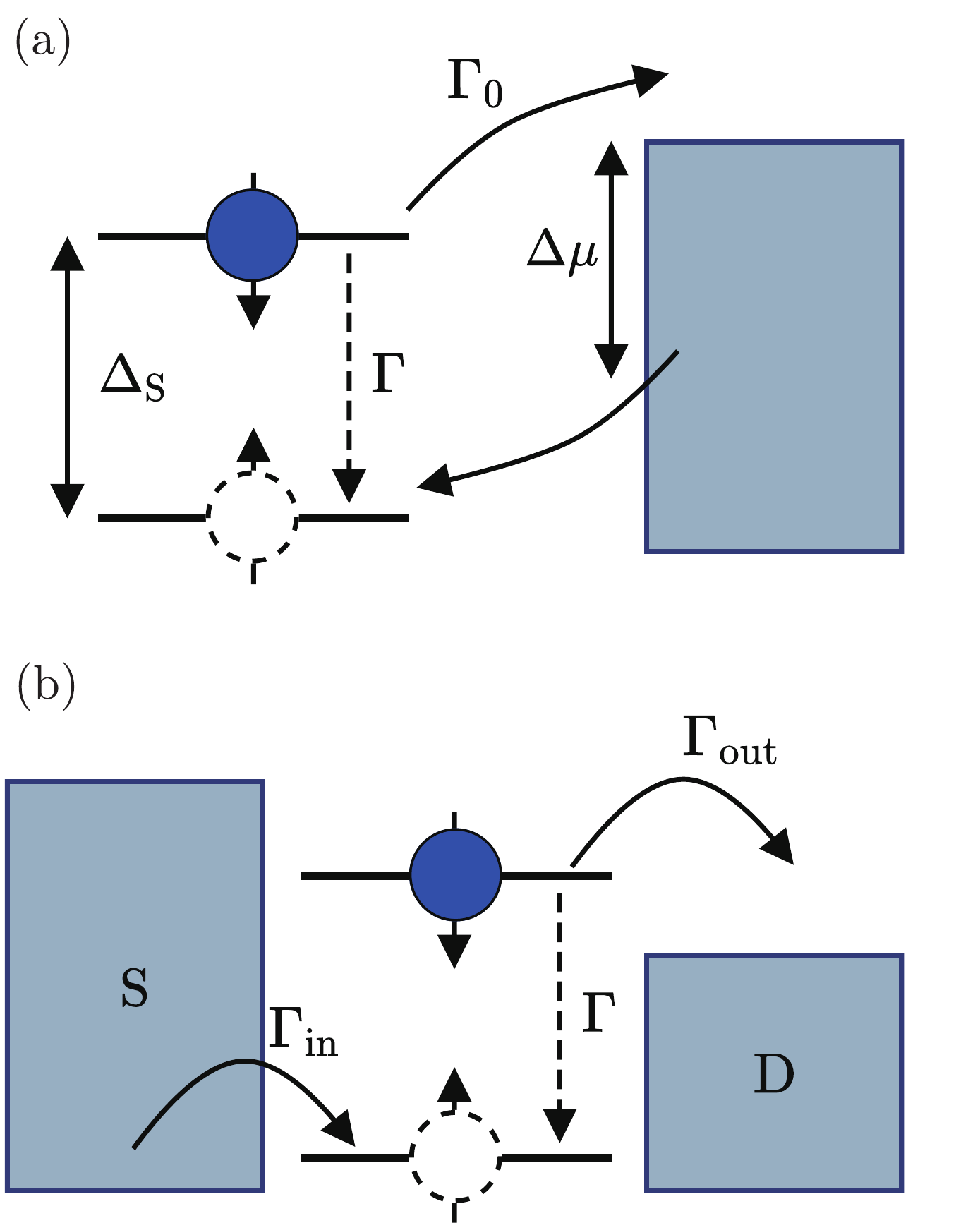}
\caption{(Color online)  (a) A cotunneling transition leading to a central-spin flip. An electron with spin down in the dot excited state tunnels into an empty state in the lead, and a spin-up electron from the lead tunnels into the dot ground state. 
Here, $\Gamma_0$ is the bare tunneling rate, and $\Delta\mu$ is the difference between the dot and lead chemical potential.
(b) An electron spin in a quantum dot in the Coulomb-blockade regime. 
The coupling with the drain D allows the electron to tunnel out of the dot only when in the excited state. Once the dot is emptied, another electron from the source S can tunnel in, with rate $\Gamma_\mathrm{in}$.
When the tunneling rate $\Gamma_\mathrm{in}$ is fast compared to other energy scales of the system, the two successive processes of tunneling out with spin down and tunneling in with spin up give rise to an effective central-spin flip with rate $\Gamma$.} \label{fig:dqd_cot}
\end{figure}
 Here, we establish the conditions for the initial master equation [Eq.~\eqref{eq:master_equation_central}, derived in Appendix \ref{app:master_equation}] in a physical context.  
 In particular, we revisit the case described in Sec.~\ref{sec:N=1} of two electron-spin qubits in a double quantum dot, where one spin serves as the ancilla and the other spin (the central spin) may undergo spin flips. 
 In this case, the universal curve depicted in Fig.~\ref{fig:universal_curve} can be realized by measuring the spin-flip time $\tau$ of the ancilla for a range of dissipation rates $\Gamma$.
When the rate $\Gamma$ can be optimized, the ancilla- and central-spin qubits can be rapidly intialized, even when the ancilla does not undergo direct spin relaxation. 

The Hamiltonian for this two-spin model, assuming an isotropic (Heisenberg) exchange ($\alpha=1$), is given by:
\begin{equation}
H_\mathrm{CS} = g_1^{*}\mu_\mathrm{B} B_1 S^1_z + g_2^{*}\mu_\mathrm{B} B_2 S^2_z + J \mathbf{S}^1\cdot\mathbf{S}^2\label{eq:Hdqd},
\end{equation} 
where $\mu_\mathrm{B}$ is the Bohr magneton, $g_1^*$ ($g_2^*$) is the g-factor of the first (second) electron, $B_1$ ($B_2$) is the local magnetic field acting on the first (second) spin, and $J$ is the Heisenberg exchange coupling. 

Equation \eqref{eq:Hdqd} maps onto the Hamiltonian given in Eq.~\eqref{eq:Hamiltonian} for:
\begin{align}
    \Delta_\mathrm{S} &= -g_1^*\mu_B B_1,\\\Delta_\mathrm{I} &= -g_2^*\mu_B B_2,\\
    \mathbf{S}&=\mathbf{S}^1,\\\mathbf{I}&=\mathbf{S}^2.
\end{align}
The master equation [Eq.~\eqref{eq:master_equation_central}] is recovered when a spin-relaxation process acts only on the central spin $\mathbf{S}=\mathbf{S}^1$. 
This situation can be realized if, e.g., only one of the dots is strongly tunnel-coupled to a lead.  
This tunnel coupling may give rise to inelastic spin-flip cotunneling transitions.  
Here, an electron in a spin excited state tunnels out into a lead, creating a virtual intermediate state. 
The electron is then replaced from the lead by an electron having the opposite spin [Fig.~\ref{fig:dqd_cot}(a)]. 
Having a double quantum dot with only one dot strongly coupled to a lead may be favorable for high-fidelity spin-to-charge conversion schemes for spin read-out \cite{studenikin2012enhanced,yang2014charge, harvey2018high}, and is therefore important in many spin-qubit devices.

For an exchange-coupled pair of electron spins (as considered here), the inelastic cotunneling process may generally describe transitions between correlated (e.g., singlet and triplet) eigenstates, rather than the single-spin states shown in Fig.~\ref{fig:dqd_cot}(a).  
However, even for a finite exchange coupling $J$, there is a regime of applicability for the process shown in Fig.~\ref{fig:dqd_cot}(a) involving a single spin flip [this leads to the master equation, Eq.~\eqref{eq:master_equation_central}].  
This regime can be reached when the level broadening $1/\tau_c$, due to a correlation time $\tau_c$, is large compared to the exchange, $J$ (the wide-band limit introduced in Appendix \ref{app:master_equation}).
The correlation time $\tau_c$ may be set by the lifetime of the intermediate state due, e.g., to a finite quasiparticle lifetime in the lead, electron-phonon coupling, or coupling to electric-field noise.  In the low-temperature limit ($k_\mathrm{B} T\lesssim \Delta_\mathrm{S}$), the spin-flip cotunneling rate $\Gamma$, within the range of applicability of a Markov approximation ($\Gamma\tau_c\ll 1$), is then given by \cite{qassemi2009stationary}:
\begin{align}
\Gamma &= \frac{\Gamma_0^2}{2\pi\Delta\mu^2}\Delta_\mathrm{S},\label{eq:cotunneling_Gamma}
\end{align}
where $\Gamma_0$ is the direct dot-lead tunnel rate, and where $\Delta\mu=\mu-(\mu_\uparrow+\mu_\downarrow)/2$, with lead chemical potential $\mu$, and dot chemical potentials $\mu_\sigma$, for spin $\sigma=\{\uparrow,\downarrow\}$. 
Importantly, the parameters $\Gamma_0$ and $\Delta\mu$ can be tuned experimentally by controlling a tunnel barrier to the lead (in the case of $\Gamma_0$) or by controlling the local dot potential (for $\Delta\mu$).  

Excitation processes [leading to spin flips $\uparrow\to\downarrow$ in Fig.~\ref{fig:dqd_cot}(a)] will be exponentially suppressed relative to the relaxation processes considered here in the low-temperature limit ($k_\mathrm{B}T<\Delta_\mathrm{S}$), provided the broadening of the lead levels $\sim 1/\tau_c$ is small compared to the spin splitting, i.e. $\Delta_\mathrm{S}\tau_c\gtrsim 1$ (this is the quasi-secular limit described in Appendix \ref{app:master_equation}).

In summary, under the following conditions, we expect the analysis here starting from the master equation [Eq.~\eqref{eq:master_equation_central}] to be valid:
\begin{eqnarray}
\Gamma\tau_c &\ll & 1,\quad \mathrm{Markov},\label{eq:MarkovConditionPhys}\\
J\tau_c &\ll & 1,\quad \mathrm{wide{-}band},\label{eq:WBConditionPhys}\\
\Delta_\mathrm{S}\tau_c & \gtrsim & 1,\quad\mathrm{quasi{-}secular}.\label{eq:QuasiSecularConditionPhys}
\end{eqnarray}

In the analysis given above, we have also neglected the effect of spin dephasing on a time scale $T_2^*$. 
We expect the effect of dephasing to be negligible provided the dephasing time is long compared to the time scale for spin flips, i.e.:
\begin{equation}
    \Gamma T_2^*\gg 1.\label{eq:relaxationCondition}
\end{equation}

To establish the experimental relevance of Fig.~\ref{fig:universal_curve}, and to show that rapid initialization to the state $\ket{\uparrow_1\uparrow_2}$ can be achieved, we now consider explicit parameters for a silicon double quantum dot.
Assuming a uniform magnetic field and equivalent $g$-factors, we set $\Delta=\Delta_\mathrm{S}-\Delta_\mathrm{I}=0$.  The singlet state of two electron spin qubits, $(\ket{\uparrow_1\downarrow_2}-\ket{\downarrow_1\uparrow_2})/\sqrt{2}$, can be rapidly prepared electrically \cite{petta2005coherent}. With this initial condition [Eq.~\eqref{eq:initial_singlet}], the optimal value for $\Gamma$ given by Eq.~\eqref{eq:tau-opt-Delta-zero} is $\hbar\Gamma_\mathrm{opt}= 2J$ and the optimal polarization time is given by $\tau_\mathrm{min}=2\hbar/J$. Assuming common experimental conditions for two spin qubits in a silicon double quantum dot, we choose $\Delta_\mathrm{S}=\Delta_\mathrm{I}=100$ $\mu$eV (corresponding to $B\simeq -1$ T for $g_1^*=g_2^*=g^*=2$ and $\hbar/\Delta_\mathrm{S}=7$ ps) and $\Delta\mu=1$ meV. 
For a direct tunneling rate $\Gamma_0=50$ $\mu \mathrm{eV}/\hbar$ (corresponding to $1/\Gamma_0= 0.013$ ns), we find the spin-flip cotunneling time $1/\Gamma= 16$ ns from Eq.~\eqref{eq:cotunneling_Gamma}. 
A comparable value for $\Gamma_0$ has been measured for a silicon double-quantum-dot device, where leakage current due to inelastic cotunneling has been observed \cite{lai2011pauli}.
At the optimal point, for $\Gamma\equiv\Gamma_\mathrm{opt}=2J/\hbar$, this gives $\hbar/J=32$ ns (corresponding to an exchange coupling $J=0.021$ $\mu$eV) and $\tau_\mathrm{min}=64$ ns. 
For these parameters, provided the correlation time $\tau_c$ satisfies $10^{-8}\;\mathrm{s} \gg \tau_\mathrm{c}\gtrsim 10^{-11}$ s, the conditions given in Eqs.~\eqref{eq:MarkovConditionPhys}-\eqref{eq:QuasiSecularConditionPhys} are satisfied.
In Si/SiGe quantum dots, the dephasing time $T_2^*$  exceeds $300$ ns \cite{maune2012coherent}, so that the condition given in Eq.~\eqref{eq:relaxationCondition} is also satisfied for the chosen parameters. The dephasing time $T_2^*$ can be much longer for isotopically purified $^{28}$Si, leading to $T_2^*> 100$ $\mu$s \cite{veldhorst2014addressable}. 
Thus, an ancilla electron-spin qubit can be rapidly initialized in 10's of nanoseconds provided the spin-flip rate $\Gamma$ is tuned to its optimal value.  

An effective spin-flip rate $\Gamma$ can also arise from a sequence of two direct tunneling processes, rather than from a two-step cotunneling process [see Fig.~\ref{fig:dqd_cot}(b)]. 
In this setup, an electron in the spin excited state ($\ket{\downarrow}$) may tunnel out to the drain ($\mathrm{D}$) with tunnel rate $\Gamma_\mathrm{out}$, leaving the dot in an empty state $\ket{0}$.  A second electron may then tunnel into either spin state ($\ket{\uparrow},\ket{\downarrow}$) with rate $\Gamma_\mathrm{in}$.  If the excited state $\ket{\downarrow}$ is occupied, the cycle continues, but eventually the spin ground state $\ket{\uparrow}$ will be loaded.  When $\Gamma_\mathrm{in}\gg\Gamma_\mathrm{out}$, the state $\ket{0}$ is only populated for a short time $\tau_c\sim 1/\Gamma_\mathrm{in}$, and the transition from $\ket{\downarrow}$ to $\ket{\uparrow}$ can be described by an effective spin-flip rate $\Gamma=\Gamma_\mathrm{out}/2$.

In addition to the example of two electron spin qubits in a double quantum dot ($N=1,I=1/2$), there are several other physical systems where the more general model ($I>1/2$) can be realized.
One such example is provided by an electron interacting with a single nuclear spin in a donor impurity coupled to leads (e.g. $\prescript{123}{}{\text{Sb}}$, for which $I=7/2$, see Ref.~\onlinecite{asaad2020coherent}).
Another possibility is to contact a spin impurity with a scanning tunneling microscope (STM) tip.  With a driven current through the impurity, this could realize Fig.~\ref{fig:dqd_cot}(b).  A similar setup has recently been used to explore the dynamics of a $^{63}$Cu or $^{65}$Cu impurity spin ($S=1/2$) in contact with a nuclear spin $I=3/2$ \cite{yang2018electrically}.
For nuclear spin $I>1/2$, in general a quadrupolar interaction must also be added to the Hamiltonian $H_\mathrm{CS}$.  This quadrupolar interaction can, however, be zero or negligible in situations of high symmetry \cite{coish2009nuclear}.
In all these examples, the conditions for the Markov, wide-band, and quasi-secular limits must be checked, to ensure that the master equation given in Eq.~\eqref{eq:master_equation_central} can be applied. These conditions are given in Eqs.~\eqref{eq:MarkovConditionPhys}-\eqref{eq:QuasiSecularConditionPhys} for $I=1/2$.  The analogous conditions for a general $I$ are given in Appendix \ref{app:master_equation}.

\section{Discussion and conclusions}
\label{sec:discussion}
In this paper, we have analyzed dissipative dynamics of a central-spin system. 
For this model, dissipation acts directly to polarize only the central spin.  Together with a coherent interaction between the central and ancilla spins, dissipation drives the ancilla spins toward a non-equilibrium (fully-polarized) steady-state.  
A set of physical conditions for the dynamical master equation describing this model [Eq.~\eqref{eq:master_equation_central}] is established in Appendix \ref{app:master_equation}.  
When these conditions are satisfied, the model will give an accurate description for some physical system.  
Outside of this regime of validity we do not expect this master equation to hold, and the steady-state will in general be different.

The central quantity of interest in this paper is the time to reach a steady state $\tau$.  
For the model given in Eq.~\eqref{eq:master_equation_central} (where central-spin flips can only occur in one direction), the dynamics can be broken up into subspaces.  
In this case, the total steady-state time can be written as a sum of steady-state times associated with independent quantum jumps [Eq.~\eqref{eq:tau-sum}, Fig.~\ref{fig:subspace_dynamics}]. 
We then find a closed-form exact analytical expression for $\tau$ [Eq.~\eqref{eq:tau_tilde}].  
This exact solution is valid over a wide range of the dissipation rate $\Gamma$.  
In particular, this analysis goes beyond the strong-dissipation (quantum-Zeno) limit (analogous to the bad-cavity limit of the Dicke model of superradiance).  
The analytical solution for $\tau$ shows a nonmonotonic behavior, with a minimum as a function of $\Gamma$.  
This minimum is described by the universal curve shown in Fig.~\ref{fig:universal_curve}. 
After scaling $\tau$ and $\Gamma$ by their optimal values, this curve is independent of the ancilla-spin length $I$ and the detuning $\Delta$, indicating the universal character of the dynamics.  
An experimental realization of the curve shown in Fig.~\ref{fig:universal_curve} would be an interesting demonstration of the interplay between coherent and dissipative dynamics.

There are many physical realizations of this model.  
One candidate, discussed in Sec.~\ref{sec:N>1}, is given by the Dicke model of superradiance (a single leaky cavity mode coupled to an ensemble of two-level systems).  
Another important example is provided by two exchange-coupled spin qubits ($N=1,I=1/2$), only one of which can undergo spin relaxation with rate $\Gamma$. 
As discussed in Sec.~\ref{sec:physical_applications}, the analysis presented here provides a strategy to optimally initialize an ancilla qubit (e.g., an electron spin in a quantum dot), even when the ancilla does not undergo direct spin relaxation.  
Physical examples for a larger ancilla-spin system ($N>1,I>1/2$) include electron spins bound to donor impurities or to adatoms interacting via the Fermi contact hyperfine interaction with a nuclear spin $I>1/2$.  
In these cases, it may be important to minimize the time to reach a fully-polarized nuclear-spin state by tuning the associated electron-spin relaxation rate $\Gamma$.

In addition to the ``natural'' physical realizations given above, it may be advantageous to design collections of qubits to follow the equation of motion derived here [Eq.~\eqref{eq:master_equation_central}].  
One example where this could be important is quantum-annealing correction \cite{pudenz2014error}. 
Quantum-annealing correction is a type of repetition code applied to quantum annealers. 
A quantum annealer is designed to approximately reach the ground state of a target Hamiltonian, by evolving from the trivial ground state of a known Hamiltonian \cite{kadowaki1998quantum}.
In practice, physical quantum annealers are subject to both coherent unitary evolution and nonunitary dissipation \cite{johnson2011quantum}.
In quantum-annealing correction, several ($N$) instances of the same Hamiltonian are simulated in parallel on independent sets of `problem' qubits.  
All equivalent problem qubits are coupled across the $N$ instances via a many-to-one interaction with an additional `penalty' qubit.  
The mean-field phase diagram for the steady state of this problem has been studied in detail \cite{matsuura2016mean,matsuura2017quantum}.
In the context of our central-spin model, the penalty qubit is provided by the central spin and each of $N=2I$ equivalent problem qubits is an ancilla spin.
For this code, it is especially important that the time scale to reach the steady state $\tau$ does not increase too quickly with the number of instances (ancillas), $N=2I$.  As we have shown in Fig.~\ref{fig:diff_scaling} and in Eq.~\eqref{eq:tau_opt_asymp}, when the dissipation rate is optimized, and for large $N=2I\gg 1$, $\tau$ does indeed increase only logarithmically with $N=2I$.  
This logarithmic dependence is a consequence of the same collective behavior that gives rise to a $\tau\propto \ln N/N$ dependence for Dicke superradiance [Eq.~\eqref{eq:DickelnNdivN}].  
It remains an interesting question for future work whether the results derived here can be used to directly improve quantum annealing protocols. 
There are several interesting questions to address before this would be possible.  One issue to address is the best method to properly engineer/tune the dissipation rate $\Gamma$ in a given physical quantum annealer. Another issue is whether the quantum-annealer steady state is robust to variations in the master equation describing the dynamics (e.g., when the conditions derived in Appendix \ref{app:master_equation} are not perfectly satisfied).

\begin{acknowledgments}
The authors thank S.~Chesi for very useful discussions.  We acknowledge funding from the Natural Sciences and Engineering Research Council (NSERC), Fonds de R\'echerche Nature et Technologies (FRQNT), and the Institut Transdisciplinaire d'Information Quantique (INTRIQ).
\end{acknowledgments}

\appendix
\section{Derivation of the central-spin master equation}
\label{app:master_equation}

The goal of this appendix is to establish the range of validity for the master equation, Eq.~\eqref{eq:master_equation_central} in the main text.  
This is an important question since Eq.~\eqref{eq:master_equation_central} does not apply, e.g., in the case of a vanishingly weak coupling to a dissipative bath, where we expect golden-rule type transitions between $H_\mathrm{CS}$ eigenstates (states that are correlated between the central and ancilla spins).  
Instead, Eq.~\eqref{eq:master_equation_central} describes quantum jumps of the central spin, independent of the ancilla.  Although there may be a less-restrictive regime for which this master equation is applicable, here we focus on sufficient conditions. 

The total Hamiltonian of the central-spin system (including spins $\mathbf{S}$ and $\mathbf{I}$) as well as an external bath (environment) is given by:
\begin{equation}
H_\mathrm{tot}=H_\mathrm{CS} + H_\mathrm{B} + H_V,
\end{equation}
where $H_\mathrm{CS}$ is the Hamiltonian of the  central spin  system defined in Eq.~\eqref{eq:Hamiltonian},
\begin{eqnarray}
H_\mathrm{CS} & = & H_S+H_I+H_{SI},\\
H_S & = & -\Delta_S S_z;\quad H_I = -\Delta_I I_z,\\
H_{SI} & = & J\left(S_zI_z+\alpha \mathbf{S}_\perp\cdot\mathbf{I}_\perp\right).
\end{eqnarray} $H_\mathrm{B}$ is the bath Hamiltonian and $H_V$ is the Hamiltonian describing the interaction between the bath and the central spin:
\begin{equation}
H_V=S_+B^\dagger+S_-B,\label{eq:V}
\end{equation}
where $B$ is an operator that acts on the bath alone. Here, we will perform a formal derivation without specifying the detailed form for the operators $B$. 

For product-state initial conditions, $\rho_\mathrm{tot}(0)=\rho(0)\otimes\rho_{\mathrm{B}}$, the Nakajima-Zwanzig generalized master equation (GME) gives an equation-of-motion for the reduced density matrix of the central-spin system after tracing out the bath, $\rho(t)=\mathrm{Tr}_\mathrm{B}\left[\rho_\mathrm{tot}(t)\right]$ \cite{fick1990quantum}:
\begin{equation}
\dot{\rho}(t)=-i\mathcal{L}_\mathrm{CS}\rho(t)-i\int_{0}^{t}dt'\Sigma(t-t')\rho(t'),\label{eq:NZME}
\end{equation}
where the action of the Liouvillian superoperator $\mathcal{L}_\alpha$ on an arbitrary operator `$\cdot$' is given by
\begin{equation}
\mathcal{L}_\alpha\cdot=[H_\alpha,\cdot],
\end{equation}
for any $\alpha=S,SI,$ etc., and the self-energy superoperator is defined by
\begin{equation}
\Sigma(t)\cdot=-i\mathrm{Tr}_{\mathrm{B}}\{\mathcal{L}_{V}\mathcal{Q}e^{-i\mathcal{L}_{\mathrm{tot}}\mathcal{Q}t}\mathcal{QL}_{V}\rho_{\mathrm{B}}\cdot\}.\label{eq:self-energy}
\end{equation}
The superoperator $\mathcal{Q}$ is the complement of another projection superoperator $\mathcal{P}$: $\mathcal{Q}=1-\mathcal{P}$ with 
\begin{equation}
\mathcal{P}\cdot=\rho_{\mathrm{B}}\mathrm{Tr}_{\mathrm{B}}\{\cdot\}.
\end{equation}

Equation \eqref{eq:NZME} is exact for the evolution of the reduced density operator $\rho(t)$, but to make analytical progress, it is common to reduce the integro-differential equation to a simpler (time-local) differential equation under a pair of approximations (the Markov and secular approximations).  These approximations take advantage of the separation in time scales for the fast bath degrees of freedom relative to the slow system evolution time (Markov approximation), or the slow system-bath interaction time relative to the time scale for rapid averaging of system observables given by the inverse system level spacing (secular approximation).  To perform these approximations, we first eliminate the system evolution with the change of variables
\begin{equation}\label{eq:FastSlow}
\rho(t) = e^{-i\mathcal{L}_\mathrm{CS}t}\tilde{\rho}(t),
\end{equation}  
and introduce the new integration variable $\tau=t-t'$:
\begin{equation}\label{eq:GMEtau}
\dot{\tilde{\rho}}(t)=-ie^{i\mathcal{L}_\mathrm{CS}t}\int_0^t d\tau \Sigma(\tau)e^{i\mathcal{L}_\mathrm{CS}\tau}e^{-i\mathcal{L}_\mathrm{CS}t}\tilde{\rho}(t-\tau).
\end{equation}

The basic assumptions of our model are that: (i) The matrix elements of $\Sigma(\tau)e^{i\mathcal{L}_\mathrm{CS}\tau}$ decay to zero sufficiently quickly (meaning the integral above converges to a finite value) over the bath correlation time $\tau_c$, and (ii) The matrix elements of $\tilde{\rho}(t)$ vary slowly up to the self-consistently determined system decay time, $t\lesssim \Gamma^{-1}$.  To simplify the presentation, we assume here that the system energy shifts (Lamb shifts) due to the bath are small compared to the self-consistently determined decay rate $\Gamma$.  If this were not the case, we could incorporate the Lamb shifts directly in the transformation of Eq.~\eqref{eq:FastSlow} (see, e.g., Chapter 13.3.2 of Ref.~\onlinecite{fick1990quantum}). Under these approximations, we set 
\begin{equation}
\tilde{\rho}(t-\tau)\simeq \tilde{\rho}(t)\quad(\Gamma\tau_c\ll 1,\,\mathrm{Markov})
\end{equation}
in the integrand of Eq.~\eqref{eq:GMEtau}. Provided we are interested in dynamics on a time scale $t\sim\Gamma^{-1}\gg\tau_c$, we extend the upper limit of integration to $t\to\infty$, giving a time-local master equation:
\begin{eqnarray}
\dot{\tilde{\rho}}(t) & = & -i\tilde{\mathcal{L}}(t)\tilde{\rho}(t),\label{eq:markov}\\
\tilde{\mathcal{L}}(t) & = & e^{i\mathcal{L}_\mathrm{CS}t}\mathcal{L}e^{-i\mathcal{L}_\mathrm{CS}t},\\
\mathcal{L} & = & \int_0^\infty d\tau \Sigma(\tau)e^{i\mathcal{L}_\mathrm{CS}\tau}.\label{eq:LDefinition}
\end{eqnarray}
Rewriting Eq.~\eqref{eq:markov} in terms of $\rho$,
\begin{equation}\label{eq:LabFrameMarkov}
\dot{\rho}(t)=-i\mathcal{L}_\mathrm{CS}\rho(t)-i\mathcal{L}\rho(t).
\end{equation}

The superoperator $\mathcal{L}$ describes energy shifts as well as excitation/relaxation of the central-spin system due to its interaction with the bath.  In the extreme limit where $\Sigma(\tau)$ decays instantly, $\tau_c\to 0$, we could take $\mathcal{L}_\mathrm{CS}\tau\to 0$ in Eq.~\eqref{eq:LDefinition}, but this could lead to an $\mathcal{L}$ that is insensitive to all spectral features of $H_\mathrm{CS}$.  In this limit, there may be no distinction between excitation/relaxation processes, and the assumption of our model (that only central-spin relaxation processes are active) would be invalid.  In the opposite limit, all spectral features of $H_\mathrm{CS}$ are relevant, and the associated dissipator would describe energy-conserving transitions between $H_\mathrm{CS}$ eigenstates, rather than flips of the central spin $\mathbf{S}$, independent of the ancilla $\mathbf{I}$.  In this paper, we consider an intermediate regime, where the relevant dissipator may be sensitive to the Zeeman term, $\mathcal{L}_z=\mathcal{L}_S+\mathcal{L}_I$, but is insensitive to the spectral features generated by $\mathcal{L}_{SI}$.  This (wide-band) limit is realized for a bandwidth (inverse correlation time $\tau_c^{-1}$) that satisfies:
\begin{equation}\label{eq:IntermediateMarkov}
||\mathcal{L}_{SI}||\tau_c\ll 1\quad \mathrm{(wide{-}band)},
\end{equation} 
where $||\cdots||$ indicates a superoperator norm, e.g., the largest eigenvalue. In the weak-coupling (Born) approximation, $\Sigma(\tau)\simeq \Sigma^{(2)}(\tau)$, only the leading-order system-bath interaction $H_V$ is taken into account, giving
\begin{equation}\label{eq:Sigma2}
\Sigma^{(2)}(\tau)=-i\mathrm{Tr}_B\mathcal{L}_Ve^{-i\left(\mathcal{L}_\mathrm{CS}+\mathcal{L}_\mathrm{B}\right)\tau}\mathcal{L}_V\rho_\mathrm{B}.
\end{equation}
We now set $\mathcal{L}_\mathrm{CS}\tau\simeq \mathcal{L}_z\tau$ in Eqs.~\eqref{eq:LDefinition}, \eqref{eq:Sigma2}, consistent with the regime given in Eq.~\eqref{eq:IntermediateMarkov}, and further note that $\left[\mathcal{L}_I,\mathcal{L}_V\right]=0$, since $H_V$ only acts on the central spin $\mathbf{S}$.  In this case, Eqs.~\eqref{eq:LDefinition} and \eqref{eq:Sigma2} give:
\begin{multline}\label{eq:CSDissipatorApprox}
\mathcal{L} \simeq -i\int_0^\infty d\tau\mathrm{Tr}_\mathrm{B}\mathcal{L}_Ve^{-i\left(\mathcal{L}_S+\mathcal{L}_\mathrm{B}\right)\tau}\mathcal{L}_V\rho_\mathrm{B}e^{i\mathcal{L}_S\tau},\\
||\mathcal{L}_{SI}||\tau_c\ll 1,\quad \Gamma\tau_c\ll 1.
\end{multline}

Under the conditions described above, the resulting super operator $\mathcal{L}$ acts only on the central spin $\mathbf{S}$.  Its action can be evaluated directly from the expression in Eq.~\eqref{eq:CSDissipatorApprox}, giving (neglecting frequency shifts):
\begin{equation}
-i\mathcal{L}\rho \simeq \Gamma_+\mathcal{D}[S_+]\rho+\Gamma_-\mathcal{D}[S_-]\rho,
\end{equation}
where 
\begin{align}
\Gamma_\pm &= \int_{-\infty}^\infty d\tau e^{\pm i\Delta_S \tau}C_\pm(\tau),\label{eq:Dissipators}\\
C_+(\tau)&=\left<B(\tau)B^\dagger\right>,\quad C_-(\tau)=\left<B^\dagger(\tau)B\right>,\label{eq:CorrelationFuncs}
\end{align}
where $\left<\cdot\right>\equiv\mathrm{Tr}_{\mathrm{B}}\left\{\cdot\rho_\mathrm{B}\right\}$.
Throughout this paper, we work in the limit where the excitation rate ($\Gamma_-$) vanishes:
\begin{equation}
\Gamma_-=0;\quad \Gamma_+=\Gamma.
\end{equation}
This limit is realized for a low-temperature environment ($k_\mathrm{B}T<\Delta_S$) provided the central-spin splitting $\Delta_S>0$ is resolvable over the correlation time of the bath:
\begin{equation}
\Delta_S\tau_c\gtrsim 1\quad (\mathrm{quasi{-}secular}).  
\end{equation}

In summary, noting that $||\mathcal{L}_{SI}||\sim JI\mathrm{max}\left(1,|\alpha|\right)$, we have the following conditions for the central-spin master equation:
\begin{eqnarray}
\Gamma\tau_c &\ll & 1,\quad \mathrm{Markov},\label{eq:MarkovCondition}\\
J I\cdot\mathrm{max}\left(1,|\alpha|\right)\tau_c & \ll & 1,\quad \mathrm{wide{-}band},\label{eq:WideBandCondition}\\
\Delta_S\tau_c & \gtrsim & 1,\quad\mathrm{quasi{-}secular}.\label{eq:QuasiSecularCondition}
\end{eqnarray}

The analysis above should be contrasted with a common weak-coupling (small $\Gamma$) and secular approximation.  This pair of approximations would lead to distinct dynamics from that considered here, where transitions go between eigenstates of $H_\mathrm{CS}$ with rates given by Fermi's golden rule. 
To evaluate these approximations, superoperator matrix elements are first evaluated with respect to a basis formed from eigenstates of $H_\mathrm{CS}$: $H_\mathrm{CS}|\mu\rangle=\epsilon_{\mu}|\mu\rangle$:
\begin{equation}
\mathcal{L}_{(\gamma\delta|\mu\nu)}=\mathrm{Tr}\{|\delta\rangle\langle\gamma|\mathcal{L}|\mu\rangle\langle\nu|\}.
\end{equation}
Further introducing $\omega_{\alpha\beta}=\epsilon_\alpha-\epsilon_\beta$:
\begin{equation}
\tilde{\mathcal{L}}_{(\gamma\delta|\mu\nu)}(t)=e^{i\left(\omega_{\gamma\delta}-\omega_{\mu\nu}\right)t}\mathcal{L}_{(\gamma\delta|\mu\nu)}.
\end{equation}

When the time scale of interest is long compared to the typical inverse system level spacing, $t\sim \Gamma^{-1}>|\omega_{\gamma\delta}-\omega_{\mu\nu}|^{-1}$, the phase factors above lead to rapid averaging of certain matrix elements, resulting in the usual secular approximation:
\begin{eqnarray}
\tilde{\mathcal{L}}_{(\gamma\delta|\mu\nu)}(t) & \simeq & \delta_{\omega_{\gamma\delta},\omega_{\mu\nu}}\mathcal{L}_{(\gamma\delta|\mu\nu)},\\
\Gamma & < & |\omega_{\gamma\delta}-\omega_{\mu\nu}|\quad(\mathrm{secular}).\label{eq:secular}
\end{eqnarray} 
Under the secular approximation [Eq.~\eqref{eq:secular}], and neglecting Lamb-shift terms, Eq.~\eqref{eq:markov} reduces to a master equation that describes quantum jumps between $H_\mathrm{CS}$ eigenstates:
\begin{equation}\label{eq:secularME}
\dot{\tilde{\rho}}(t)=\sum_{\mu\nu}\Gamma_{\mu\to\nu}\mathcal{D}[|\nu\rangle\langle\mu|]\tilde{\rho}(t).
\end{equation}
For such a master equation, $H_\mathrm{CS}$ eigenstates are steady-state solutions.  This is qualitatively distinct from the behavior of central-spin master equation, Eq.~\eqref{eq:master_equation_central} in the main text.  

The secular master equation, Eq.~\eqref{eq:secularME}, is valid for a sufficiently weak $\Gamma$ [Eq.~\eqref{eq:secular}], but will not generally be valid when Eq.~\eqref{eq:secular} is not satisfied.  In this paper, we are especially interested in situations where the optimal central-spin flip rate $\Gamma\simeq \Gamma_\mathrm{opt}$ is comparable to (or larger than) the scale of $||\mathcal{L}_\mathrm{CS}||$, where the equilibration time is minimized.  In this limit, it is impossible to apply the common secular approximation [Eq.~\eqref{eq:secular}], leading instead to the less restrictive conditions given in Eqs.~\eqref{eq:MarkovCondition}, \eqref{eq:WideBandCondition}, \eqref{eq:QuasiSecularCondition}.

\bibliographystyle{apsrev4-2}
\bibliography{css-bibliography}

\begin{thebibliography}{54}%
\makeatletter
\providecommand \@ifxundefined [1]{%
 \@ifx{#1\undefined}
}%
\providecommand \@ifnum [1]{%
 \ifnum #1\expandafter \@firstoftwo
 \else \expandafter \@secondoftwo
 \fi
}%
\providecommand \@ifx [1]{%
 \ifx #1\expandafter \@firstoftwo
 \else \expandafter \@secondoftwo
 \fi
}%
\providecommand \natexlab [1]{#1}%
\providecommand \enquote  [1]{``#1''}%
\providecommand \bibnamefont  [1]{#1}%
\providecommand \bibfnamefont [1]{#1}%
\providecommand \citenamefont [1]{#1}%
\providecommand \href@noop [0]{\@secondoftwo}%
\providecommand \href [0]{\begingroup \@sanitize@url \@href}%
\providecommand \@href[1]{\@@startlink{#1}\@@href}%
\providecommand \@@href[1]{\endgroup#1\@@endlink}%
\providecommand \@sanitize@url [0]{\catcode `\\12\catcode `\$12\catcode
  `\&12\catcode `\#12\catcode `\^12\catcode `\_12\catcode `\%12\relax}%
\providecommand \@@startlink[1]{}%
\providecommand \@@endlink[0]{}%
\providecommand \url  [0]{\begingroup\@sanitize@url \@url }%
\providecommand \@url [1]{\endgroup\@href {#1}{\urlprefix }}%
\providecommand \urlprefix  [0]{URL }%
\providecommand \Eprint [0]{\href }%
\providecommand \doibase [0]{https://doi.org/}%
\providecommand \selectlanguage [0]{\@gobble}%
\providecommand \bibinfo  [0]{\@secondoftwo}%
\providecommand \bibfield  [0]{\@secondoftwo}%
\providecommand \translation [1]{[#1]}%
\providecommand \BibitemOpen [0]{}%
\providecommand \bibitemStop [0]{}%
\providecommand \bibitemNoStop [0]{.\EOS\space}%
\providecommand \EOS [0]{\spacefactor3000\relax}%
\providecommand \BibitemShut  [1]{\csname bibitem#1\endcsname}%
\let\auto@bib@innerbib\@empty
\bibitem [{\citenamefont {Gordon}\ \emph {et~al.}(2009)\citenamefont {Gordon},
  \citenamefont {Bensky}, \citenamefont {Gelbwaser-Klimovsky}, \citenamefont
  {Rao}, \citenamefont {Erez},\ and\ \citenamefont
  {Kurizki}}]{gordon2009cooling}%
  \BibitemOpen
  \bibfield  {author} {\bibinfo {author} {\bibfnamefont {G.}~\bibnamefont
  {Gordon}}, \bibinfo {author} {\bibfnamefont {G.}~\bibnamefont {Bensky}},
  \bibinfo {author} {\bibfnamefont {D.}~\bibnamefont {Gelbwaser-Klimovsky}},
  \bibinfo {author} {\bibfnamefont {D.~B.}\ \bibnamefont {Rao}}, \bibinfo
  {author} {\bibfnamefont {N.}~\bibnamefont {Erez}},\ and\ \bibinfo {author}
  {\bibfnamefont {G.}~\bibnamefont {Kurizki}},\ }\href@noop {} {\bibfield
  {journal} {\bibinfo  {journal} {New J. Phys.}\ }\textbf {\bibinfo {volume}
  {11}},\ \bibinfo {pages} {123025} (\bibinfo {year} {2009})}\BibitemShut
  {NoStop}%
\bibitem [{\citenamefont {{\'A}lvarez}\ \emph {et~al.}(2010)\citenamefont
  {{\'A}lvarez}, \citenamefont {Rao}, \citenamefont {Frydman},\ and\
  \citenamefont {Kurizki}}]{alvarez2010zeno}%
  \BibitemOpen
  \bibfield  {author} {\bibinfo {author} {\bibfnamefont {G.~A.}\ \bibnamefont
  {{\'A}lvarez}}, \bibinfo {author} {\bibfnamefont {D.~D.~B.}\ \bibnamefont
  {Rao}}, \bibinfo {author} {\bibfnamefont {L.}~\bibnamefont {Frydman}},\ and\
  \bibinfo {author} {\bibfnamefont {G.}~\bibnamefont {Kurizki}},\ }\href@noop
  {} {\bibfield  {journal} {\bibinfo  {journal} {\prl}\ }\textbf {\bibinfo
  {volume} {105}},\ \bibinfo {pages} {160401} (\bibinfo {year}
  {2010})}\BibitemShut {NoStop}%
\bibitem [{\citenamefont {Misra}\ and\ \citenamefont
  {Sudarshan}(1977)}]{misra1977zeno}%
  \BibitemOpen
  \bibfield  {author} {\bibinfo {author} {\bibfnamefont {B.}~\bibnamefont
  {Misra}}\ and\ \bibinfo {author} {\bibfnamefont {E.~C.~G.}\ \bibnamefont
  {Sudarshan}},\ }\href@noop {} {\bibfield  {journal} {\bibinfo  {journal}
  {J.~Math.~Phys.~}\ }\textbf {\bibinfo {volume} {18}},\ \bibinfo {pages} {756}
  (\bibinfo {year} {1977})}\BibitemShut {NoStop}%
\bibitem [{\citenamefont {Peres}(1980)}]{peres1980zeno}%
  \BibitemOpen
  \bibfield  {author} {\bibinfo {author} {\bibfnamefont {A.}~\bibnamefont
  {Peres}},\ }\href@noop {} {\bibfield  {journal} {\bibinfo  {journal}
  {Am.~J.~Phys.~}\ }\textbf {\bibinfo {volume} {48}},\ \bibinfo {pages} {931}
  (\bibinfo {year} {1980})}\BibitemShut {NoStop}%
\bibitem [{\citenamefont {Gambetta}\ \emph {et~al.}(2008)\citenamefont
  {Gambetta}, \citenamefont {Blais}, \citenamefont {Boissonneault},
  \citenamefont {Houck}, \citenamefont {Schuster},\ and\ \citenamefont
  {Girvin}}]{gambetta2008quantum}%
  \BibitemOpen
  \bibfield  {author} {\bibinfo {author} {\bibfnamefont {J.}~\bibnamefont
  {Gambetta}}, \bibinfo {author} {\bibfnamefont {A.}~\bibnamefont {Blais}},
  \bibinfo {author} {\bibfnamefont {M.}~\bibnamefont {Boissonneault}}, \bibinfo
  {author} {\bibfnamefont {A.~A.}\ \bibnamefont {Houck}}, \bibinfo {author}
  {\bibfnamefont {D.~I.}\ \bibnamefont {Schuster}},\ and\ \bibinfo {author}
  {\bibfnamefont {S.~M.}\ \bibnamefont {Girvin}},\ }\href@noop {} {\bibfield
  {journal} {\bibinfo  {journal} {\pra}\ }\textbf {\bibinfo {volume} {77}},\
  \bibinfo {pages} {012112} (\bibinfo {year} {2008})}\BibitemShut {NoStop}%
\bibitem [{\citenamefont {Bernu}\ \emph {et~al.}(2008)\citenamefont {Bernu},
  \citenamefont {Del{\'e}glise}, \citenamefont {Sayrin}, \citenamefont {Kuhr},
  \citenamefont {Dotsenko}, \citenamefont {Brune}, \citenamefont {Raimond},\
  and\ \citenamefont {Haroche}}]{bernu2008freezing}%
  \BibitemOpen
  \bibfield  {author} {\bibinfo {author} {\bibfnamefont {J.}~\bibnamefont
  {Bernu}}, \bibinfo {author} {\bibfnamefont {S.}~\bibnamefont
  {Del{\'e}glise}}, \bibinfo {author} {\bibfnamefont {C.}~\bibnamefont
  {Sayrin}}, \bibinfo {author} {\bibfnamefont {S.}~\bibnamefont {Kuhr}},
  \bibinfo {author} {\bibfnamefont {I.}~\bibnamefont {Dotsenko}}, \bibinfo
  {author} {\bibfnamefont {M.}~\bibnamefont {Brune}}, \bibinfo {author}
  {\bibfnamefont {J.-M.}\ \bibnamefont {Raimond}},\ and\ \bibinfo {author}
  {\bibfnamefont {S.}~\bibnamefont {Haroche}},\ }\href@noop {} {\bibfield
  {journal} {\bibinfo  {journal} {\prl}\ }\textbf {\bibinfo {volume} {101}},\
  \bibinfo {pages} {180402} (\bibinfo {year} {2008})}\BibitemShut {NoStop}%
\bibitem [{\citenamefont {Raimond}\ \emph {et~al.}(2012)\citenamefont
  {Raimond}, \citenamefont {Facchi}, \citenamefont {Peaudecerf}, \citenamefont
  {Pascazio}, \citenamefont {Sayrin}, \citenamefont {Dotsenko}, \citenamefont
  {Gleyzes}, \citenamefont {Brune},\ and\ \citenamefont
  {Haroche}}]{raimond2012quantum}%
  \BibitemOpen
  \bibfield  {author} {\bibinfo {author} {\bibfnamefont {J.-M.}\ \bibnamefont
  {Raimond}}, \bibinfo {author} {\bibfnamefont {P.}~\bibnamefont {Facchi}},
  \bibinfo {author} {\bibfnamefont {B.}~\bibnamefont {Peaudecerf}}, \bibinfo
  {author} {\bibfnamefont {S.}~\bibnamefont {Pascazio}}, \bibinfo {author}
  {\bibfnamefont {C.}~\bibnamefont {Sayrin}}, \bibinfo {author} {\bibfnamefont
  {I.}~\bibnamefont {Dotsenko}}, \bibinfo {author} {\bibfnamefont
  {S.}~\bibnamefont {Gleyzes}}, \bibinfo {author} {\bibfnamefont
  {M.}~\bibnamefont {Brune}},\ and\ \bibinfo {author} {\bibfnamefont
  {S.}~\bibnamefont {Haroche}},\ }\href@noop {} {\bibfield  {journal} {\bibinfo
   {journal} {\pra}\ }\textbf {\bibinfo {volume} {86}},\ \bibinfo {pages}
  {032120} (\bibinfo {year} {2012})}\BibitemShut {NoStop}%
\bibitem [{\citenamefont {Segal}\ and\ \citenamefont
  {Reichman}(2007)}]{segal2007zeno}%
  \BibitemOpen
  \bibfield  {author} {\bibinfo {author} {\bibfnamefont {D.}~\bibnamefont
  {Segal}}\ and\ \bibinfo {author} {\bibfnamefont {D.~R.}\ \bibnamefont
  {Reichman}},\ }\href@noop {} {\bibfield  {journal} {\bibinfo  {journal}
  {\pra}\ }\textbf {\bibinfo {volume} {76}},\ \bibinfo {pages} {012109}
  (\bibinfo {year} {2007})}\BibitemShut {NoStop}%
\bibitem [{\citenamefont {Watson}\ \emph {et~al.}(2018)\citenamefont {Watson},
  \citenamefont {Philips}, \citenamefont {Kawakami}, \citenamefont {Ward},
  \citenamefont {Scarlino}, \citenamefont {Veldhorst}, \citenamefont {Savage},
  \citenamefont {Lagally}, \citenamefont {Friesen}, \citenamefont
  {Coppersmith}, \citenamefont {Erikson},\ and\ \citenamefont
  {Vandersypen}}]{watson2018programmable}%
  \BibitemOpen
  \bibfield  {author} {\bibinfo {author} {\bibfnamefont {T.~F.}\ \bibnamefont
  {Watson}}, \bibinfo {author} {\bibfnamefont {S.~G.~J.}\ \bibnamefont
  {Philips}}, \bibinfo {author} {\bibfnamefont {E.}~\bibnamefont {Kawakami}},
  \bibinfo {author} {\bibfnamefont {D.~R.}\ \bibnamefont {Ward}}, \bibinfo
  {author} {\bibfnamefont {P.}~\bibnamefont {Scarlino}}, \bibinfo {author}
  {\bibfnamefont {M.}~\bibnamefont {Veldhorst}}, \bibinfo {author}
  {\bibfnamefont {D.~E.}\ \bibnamefont {Savage}}, \bibinfo {author}
  {\bibfnamefont {M.~G.}\ \bibnamefont {Lagally}}, \bibinfo {author}
  {\bibfnamefont {M.}~\bibnamefont {Friesen}}, \bibinfo {author} {\bibfnamefont
  {S.~N.}\ \bibnamefont {Coppersmith}}, \bibinfo {author} {\bibfnamefont
  {M.~A.}\ \bibnamefont {Erikson}},\ and\ \bibinfo {author} {\bibfnamefont
  {L.~M.~K.}\ \bibnamefont {Vandersypen}},\ }\href@noop {} {\bibfield
  {journal} {\bibinfo  {journal} {Nature}\ }\textbf {\bibinfo {volume} {555}},\
  \bibinfo {pages} {633} (\bibinfo {year} {2018})}\BibitemShut {NoStop}%
\bibitem [{\citenamefont {Zajac}\ \emph {et~al.}(2018)\citenamefont {Zajac},
  \citenamefont {Sigillito}, \citenamefont {Russ}, \citenamefont {Borjans},
  \citenamefont {Taylor}, \citenamefont {Burkard},\ and\ \citenamefont
  {Petta}}]{zajac2018resonantly}%
  \BibitemOpen
  \bibfield  {author} {\bibinfo {author} {\bibfnamefont {D.~M.}\ \bibnamefont
  {Zajac}}, \bibinfo {author} {\bibfnamefont {A.~J.}\ \bibnamefont
  {Sigillito}}, \bibinfo {author} {\bibfnamefont {M.}~\bibnamefont {Russ}},
  \bibinfo {author} {\bibfnamefont {F.}~\bibnamefont {Borjans}}, \bibinfo
  {author} {\bibfnamefont {J.~M.}\ \bibnamefont {Taylor}}, \bibinfo {author}
  {\bibfnamefont {G.}~\bibnamefont {Burkard}},\ and\ \bibinfo {author}
  {\bibfnamefont {J.~R.}\ \bibnamefont {Petta}},\ }\href@noop {} {\bibfield
  {journal} {\bibinfo  {journal} {Science}\ }\textbf {\bibinfo {volume}
  {359}},\ \bibinfo {pages} {439} (\bibinfo {year} {2018})}\BibitemShut
  {NoStop}%
\bibitem [{\citenamefont {B{\"u}ch}\ \emph {et~al.}(2013)\citenamefont
  {B{\"u}ch}, \citenamefont {Mahapatra}, \citenamefont {Rahman}, \citenamefont
  {Morello},\ and\ \citenamefont {Simmons}}]{buch2013spin}%
  \BibitemOpen
  \bibfield  {author} {\bibinfo {author} {\bibfnamefont {H.}~\bibnamefont
  {B{\"u}ch}}, \bibinfo {author} {\bibfnamefont {S.}~\bibnamefont {Mahapatra}},
  \bibinfo {author} {\bibfnamefont {R.}~\bibnamefont {Rahman}}, \bibinfo
  {author} {\bibfnamefont {A.}~\bibnamefont {Morello}},\ and\ \bibinfo {author}
  {\bibfnamefont {M.~Y.}\ \bibnamefont {Simmons}},\ }\href@noop {} {\bibfield
  {journal} {\bibinfo  {journal} {Nat.~Commun.~}\ }\textbf {\bibinfo {volume}
  {4}},\ \bibinfo {pages} {2017} (\bibinfo {year} {2013})}\BibitemShut
  {NoStop}%
\bibitem [{\citenamefont {Franke}\ \emph {et~al.}(2016)\citenamefont {Franke},
  \citenamefont {Pfl{\"u}ger}, \citenamefont {Mortemousque}, \citenamefont
  {Itoh},\ and\ \citenamefont {Brandt}}]{franke2016quadrupolar}%
  \BibitemOpen
  \bibfield  {author} {\bibinfo {author} {\bibfnamefont {D.~P.}\ \bibnamefont
  {Franke}}, \bibinfo {author} {\bibfnamefont {M.~P.~D.}\ \bibnamefont
  {Pfl{\"u}ger}}, \bibinfo {author} {\bibfnamefont {P.-A.}\ \bibnamefont
  {Mortemousque}}, \bibinfo {author} {\bibfnamefont {K.~M.}\ \bibnamefont
  {Itoh}},\ and\ \bibinfo {author} {\bibfnamefont {M.~S.}\ \bibnamefont
  {Brandt}},\ }\href@noop {} {\bibfield  {journal} {\bibinfo  {journal} {\prb}\
  }\textbf {\bibinfo {volume} {93}},\ \bibinfo {pages} {161303(R)} (\bibinfo
  {year} {2016})}\BibitemShut {NoStop}%
\bibitem [{\citenamefont {Schenkel}\ \emph {et~al.}(2006)\citenamefont
  {Schenkel}, \citenamefont {Liddle}, \citenamefont {Persaud}, \citenamefont
  {Tyryshkin}, \citenamefont {Lyon}, \citenamefont {De~Sousa}, \citenamefont
  {Whaley}, \citenamefont {Bokor}, \citenamefont {Shangkuan},\ and\
  \citenamefont {Chakarov}}]{schenkel2006electrical}%
  \BibitemOpen
  \bibfield  {author} {\bibinfo {author} {\bibfnamefont {T.}~\bibnamefont
  {Schenkel}}, \bibinfo {author} {\bibfnamefont {J.~A.}\ \bibnamefont
  {Liddle}}, \bibinfo {author} {\bibfnamefont {A.}~\bibnamefont {Persaud}},
  \bibinfo {author} {\bibfnamefont {A.~M.}\ \bibnamefont {Tyryshkin}}, \bibinfo
  {author} {\bibfnamefont {S.~A.}\ \bibnamefont {Lyon}}, \bibinfo {author}
  {\bibfnamefont {R.}~\bibnamefont {De~Sousa}}, \bibinfo {author}
  {\bibfnamefont {K.~B.}\ \bibnamefont {Whaley}}, \bibinfo {author}
  {\bibfnamefont {J.}~\bibnamefont {Bokor}}, \bibinfo {author} {\bibfnamefont
  {J.}~\bibnamefont {Shangkuan}},\ and\ \bibinfo {author} {\bibfnamefont
  {I.}~\bibnamefont {Chakarov}},\ }\href@noop {} {\bibfield  {journal}
  {\bibinfo  {journal} {Appl.~Phys.~Lett.~}\ }\textbf {\bibinfo {volume}
  {88}},\ \bibinfo {pages} {112101} (\bibinfo {year} {2006})}\BibitemShut
  {NoStop}%
\bibitem [{\citenamefont {Muhonen}\ \emph {et~al.}(2018)\citenamefont
  {Muhonen}, \citenamefont {Dehollain}, \citenamefont {Laucht}, \citenamefont
  {Simmons}, \citenamefont {Kalra}, \citenamefont {Hudson}, \citenamefont
  {Dzurak}, \citenamefont {Morello}, \citenamefont {Jamieson}, \citenamefont
  {McCallum} \emph {et~al.}}]{muhonen2018coherent}%
  \BibitemOpen
  \bibfield  {author} {\bibinfo {author} {\bibfnamefont {J.~T.}\ \bibnamefont
  {Muhonen}}, \bibinfo {author} {\bibfnamefont {J.~P.}\ \bibnamefont
  {Dehollain}}, \bibinfo {author} {\bibfnamefont {A.}~\bibnamefont {Laucht}},
  \bibinfo {author} {\bibfnamefont {S.}~\bibnamefont {Simmons}}, \bibinfo
  {author} {\bibfnamefont {R.}~\bibnamefont {Kalra}}, \bibinfo {author}
  {\bibfnamefont {F.~E.}\ \bibnamefont {Hudson}}, \bibinfo {author}
  {\bibfnamefont {A.~S.}\ \bibnamefont {Dzurak}}, \bibinfo {author}
  {\bibfnamefont {A.}~\bibnamefont {Morello}}, \bibinfo {author} {\bibfnamefont
  {D.~N.}\ \bibnamefont {Jamieson}}, \bibinfo {author} {\bibfnamefont {J.~C.}\
  \bibnamefont {McCallum}}, \emph {et~al.},\ }\href@noop {} {\bibfield
  {journal} {\bibinfo  {journal} {\prb}\ }\textbf {\bibinfo {volume} {98}},\
  \bibinfo {pages} {155201} (\bibinfo {year} {2018})}\BibitemShut {NoStop}%
\bibitem [{\citenamefont {Chekhovich}\ \emph {et~al.}(2013)\citenamefont
  {Chekhovich}, \citenamefont {Makhonin}, \citenamefont {Tartakovskii},
  \citenamefont {Yacoby}, \citenamefont {Bluhm}, \citenamefont {Nowack},\ and\
  \citenamefont {Vandersypen}}]{chekhovich2013nuclear}%
  \BibitemOpen
  \bibfield  {author} {\bibinfo {author} {\bibfnamefont {E.~A.}\ \bibnamefont
  {Chekhovich}}, \bibinfo {author} {\bibfnamefont {M.~N.}\ \bibnamefont
  {Makhonin}}, \bibinfo {author} {\bibfnamefont {A.~I.}\ \bibnamefont
  {Tartakovskii}}, \bibinfo {author} {\bibfnamefont {A.}~\bibnamefont
  {Yacoby}}, \bibinfo {author} {\bibfnamefont {H.}~\bibnamefont {Bluhm}},
  \bibinfo {author} {\bibfnamefont {K.~C.}\ \bibnamefont {Nowack}},\ and\
  \bibinfo {author} {\bibfnamefont {L.~M.~K.}\ \bibnamefont {Vandersypen}},\
  }\href@noop {} {\bibfield  {journal} {\bibinfo  {journal} {Nat.~Mater.~}\
  }\textbf {\bibinfo {volume} {12}},\ \bibinfo {pages} {494} (\bibinfo {year}
  {2013})}\BibitemShut {NoStop}%
\bibitem [{\citenamefont {Coish}\ and\ \citenamefont
  {Baugh}(2009)}]{coish2009nuclear}%
  \BibitemOpen
  \bibfield  {author} {\bibinfo {author} {\bibfnamefont {W.~A.}\ \bibnamefont
  {Coish}}\ and\ \bibinfo {author} {\bibfnamefont {J.}~\bibnamefont {Baugh}},\
  }\href@noop {} {\bibfield  {journal} {\bibinfo  {journal} {Phys.~Status
  Solidi B}\ }\textbf {\bibinfo {volume} {246}},\ \bibinfo {pages} {2203}
  (\bibinfo {year} {2009})}\BibitemShut {NoStop}%
\bibitem [{\citenamefont {Kessler}\ \emph {et~al.}(2012)\citenamefont
  {Kessler}, \citenamefont {Giedke}, \citenamefont {Imamoglu}, \citenamefont
  {Yelin}, \citenamefont {Lukin},\ and\ \citenamefont
  {Cirac}}]{kessler2012dissipative}%
  \BibitemOpen
  \bibfield  {author} {\bibinfo {author} {\bibfnamefont {E.~M.}\ \bibnamefont
  {Kessler}}, \bibinfo {author} {\bibfnamefont {G.}~\bibnamefont {Giedke}},
  \bibinfo {author} {\bibfnamefont {A.}~\bibnamefont {Imamoglu}}, \bibinfo
  {author} {\bibfnamefont {S.~F.}\ \bibnamefont {Yelin}}, \bibinfo {author}
  {\bibfnamefont {M.~D.}\ \bibnamefont {Lukin}},\ and\ \bibinfo {author}
  {\bibfnamefont {J.~I.}\ \bibnamefont {Cirac}},\ }\href@noop {} {\bibfield
  {journal} {\bibinfo  {journal} {\pra}\ }\textbf {\bibinfo {volume} {86}},\
  \bibinfo {pages} {012116} (\bibinfo {year} {2012})}\BibitemShut {NoStop}%
\bibitem [{\citenamefont {Fuchs}\ \emph {et~al.}(2015)\citenamefont {Fuchs},
  \citenamefont {Krau{\ss}}, \citenamefont {Hetterich},\ and\ \citenamefont
  {Trauzettel}}]{fuchs2015thermal}%
  \BibitemOpen
  \bibfield  {author} {\bibinfo {author} {\bibfnamefont {M.}~\bibnamefont
  {Fuchs}}, \bibinfo {author} {\bibfnamefont {F.}~\bibnamefont {Krau{\ss}}},
  \bibinfo {author} {\bibfnamefont {D.}~\bibnamefont {Hetterich}},\ and\
  \bibinfo {author} {\bibfnamefont {B.}~\bibnamefont {Trauzettel}},\
  }\href@noop {} {\bibfield  {journal} {\bibinfo  {journal} {\prb}\ }\textbf
  {\bibinfo {volume} {92}},\ \bibinfo {pages} {035310} (\bibinfo {year}
  {2015})}\BibitemShut {NoStop}%
\bibitem [{\citenamefont {Rudner}\ and\ \citenamefont
  {Levitov}(2010)}]{rudner2010phase}%
  \BibitemOpen
  \bibfield  {author} {\bibinfo {author} {\bibfnamefont {M.~S.}\ \bibnamefont
  {Rudner}}\ and\ \bibinfo {author} {\bibfnamefont {L.~S.}\ \bibnamefont
  {Levitov}},\ }\href@noop {} {\bibfield  {journal} {\bibinfo  {journal}
  {\prb}\ }\textbf {\bibinfo {volume} {82}},\ \bibinfo {pages} {155418}
  (\bibinfo {year} {2010})}\BibitemShut {NoStop}%
\bibitem [{\citenamefont {He}\ \emph {et~al.}(2019{\natexlab{a}})\citenamefont
  {He}, \citenamefont {Chesi}, \citenamefont {Lin},\ and\ \citenamefont
  {Guan}}]{he2019exact}%
  \BibitemOpen
  \bibfield  {author} {\bibinfo {author} {\bibfnamefont {W.-B.}\ \bibnamefont
  {He}}, \bibinfo {author} {\bibfnamefont {S.}~\bibnamefont {Chesi}}, \bibinfo
  {author} {\bibfnamefont {H.-Q.}\ \bibnamefont {Lin}},\ and\ \bibinfo {author}
  {\bibfnamefont {X.-W.}\ \bibnamefont {Guan}},\ }\href@noop {} {\bibfield
  {journal} {\bibinfo  {journal} {\prb}\ }\textbf {\bibinfo {volume} {99}},\
  \bibinfo {pages} {174308} (\bibinfo {year} {2019}{\natexlab{a}})}\BibitemShut
  {NoStop}%
\bibitem [{\citenamefont {Dooley}\ \emph {et~al.}(2013)\citenamefont {Dooley},
  \citenamefont {McCrossan}, \citenamefont {Harland}, \citenamefont {Everitt},\
  and\ \citenamefont {Spiller}}]{dooley2013collapse}%
  \BibitemOpen
  \bibfield  {author} {\bibinfo {author} {\bibfnamefont {S.}~\bibnamefont
  {Dooley}}, \bibinfo {author} {\bibfnamefont {F.}~\bibnamefont {McCrossan}},
  \bibinfo {author} {\bibfnamefont {D.}~\bibnamefont {Harland}}, \bibinfo
  {author} {\bibfnamefont {M.~J.}\ \bibnamefont {Everitt}},\ and\ \bibinfo
  {author} {\bibfnamefont {T.~P.}\ \bibnamefont {Spiller}},\ }\href@noop {}
  {\bibfield  {journal} {\bibinfo  {journal} {\pra}\ }\textbf {\bibinfo
  {volume} {87}},\ \bibinfo {pages} {052323} (\bibinfo {year}
  {2013})}\BibitemShut {NoStop}%
\bibitem [{\citenamefont {Danon}\ and\ \citenamefont
  {Nazarov}(2008)}]{danon2008tuning}%
  \BibitemOpen
  \bibfield  {author} {\bibinfo {author} {\bibfnamefont {J.}~\bibnamefont
  {Danon}}\ and\ \bibinfo {author} {\bibfnamefont {Y.~V.}\ \bibnamefont
  {Nazarov}},\ }\href {https://doi.org/10.1103/PhysRevLett.100.056603}
  {\bibfield  {journal} {\bibinfo  {journal} {\prl}\ }\textbf {\bibinfo
  {volume} {100}},\ \bibinfo {pages} {056603} (\bibinfo {year}
  {2008})}\BibitemShut {NoStop}%
\bibitem [{\citenamefont {Danon}\ and\ \citenamefont
  {Nazarov}(2011)}]{danon2011nuclear}%
  \BibitemOpen
  \bibfield  {author} {\bibinfo {author} {\bibfnamefont {J.}~\bibnamefont
  {Danon}}\ and\ \bibinfo {author} {\bibfnamefont {Y.~V.}\ \bibnamefont
  {Nazarov}},\ }\href {https://doi.org/10.1103/PhysRevB.83.245306} {\bibfield
  {journal} {\bibinfo  {journal} {\prb}\ }\textbf {\bibinfo {volume} {83}},\
  \bibinfo {pages} {245306} (\bibinfo {year} {2011})}\BibitemShut {NoStop}%
\bibitem [{\citenamefont {Gullans}\ \emph {et~al.}(2010)\citenamefont
  {Gullans}, \citenamefont {Krich}, \citenamefont {Taylor}, \citenamefont
  {Bluhm}, \citenamefont {Halperin}, \citenamefont {Marcus}, \citenamefont
  {Stopa}, \citenamefont {Yacoby},\ and\ \citenamefont
  {Lukin}}]{gullans2010dynamic}%
  \BibitemOpen
  \bibfield  {author} {\bibinfo {author} {\bibfnamefont {M.}~\bibnamefont
  {Gullans}}, \bibinfo {author} {\bibfnamefont {J.~J.}\ \bibnamefont {Krich}},
  \bibinfo {author} {\bibfnamefont {J.~M.}\ \bibnamefont {Taylor}}, \bibinfo
  {author} {\bibfnamefont {H.}~\bibnamefont {Bluhm}}, \bibinfo {author}
  {\bibfnamefont {B.~I.}\ \bibnamefont {Halperin}}, \bibinfo {author}
  {\bibfnamefont {C.~M.}\ \bibnamefont {Marcus}}, \bibinfo {author}
  {\bibfnamefont {M.}~\bibnamefont {Stopa}}, \bibinfo {author} {\bibfnamefont
  {A.}~\bibnamefont {Yacoby}},\ and\ \bibinfo {author} {\bibfnamefont {M.~D.}\
  \bibnamefont {Lukin}},\ }\href@noop {} {\bibfield  {journal} {\bibinfo
  {journal} {\prl}\ }\textbf {\bibinfo {volume} {104}},\ \bibinfo {pages}
  {226807} (\bibinfo {year} {2010})}\BibitemShut {NoStop}%
\bibitem [{\citenamefont {Gullans}\ \emph {et~al.}(2013)\citenamefont
  {Gullans}, \citenamefont {Krich}, \citenamefont {Taylor}, \citenamefont
  {Halperin},\ and\ \citenamefont {Lukin}}]{gullans2013preparation}%
  \BibitemOpen
  \bibfield  {author} {\bibinfo {author} {\bibfnamefont {M.}~\bibnamefont
  {Gullans}}, \bibinfo {author} {\bibfnamefont {J.~J.}\ \bibnamefont {Krich}},
  \bibinfo {author} {\bibfnamefont {J.~M.}\ \bibnamefont {Taylor}}, \bibinfo
  {author} {\bibfnamefont {B.~I.}\ \bibnamefont {Halperin}},\ and\ \bibinfo
  {author} {\bibfnamefont {M.~D.}\ \bibnamefont {Lukin}},\ }\href@noop {}
  {\bibfield  {journal} {\bibinfo  {journal} {\prb}\ }\textbf {\bibinfo
  {volume} {88}},\ \bibinfo {pages} {035309} (\bibinfo {year}
  {2013})}\BibitemShut {NoStop}%
\bibitem [{\citenamefont {Neder}\ \emph {et~al.}(2014)\citenamefont {Neder},
  \citenamefont {Rudner},\ and\ \citenamefont {Halperin}}]{neder2014theory}%
  \BibitemOpen
  \bibfield  {author} {\bibinfo {author} {\bibfnamefont {I.}~\bibnamefont
  {Neder}}, \bibinfo {author} {\bibfnamefont {M.~S.}\ \bibnamefont {Rudner}},\
  and\ \bibinfo {author} {\bibfnamefont {B.~I.}\ \bibnamefont {Halperin}},\
  }\href@noop {} {\bibfield  {journal} {\bibinfo  {journal} {\prb}\ }\textbf
  {\bibinfo {volume} {89}},\ \bibinfo {pages} {085403} (\bibinfo {year}
  {2014})}\BibitemShut {NoStop}%
\bibitem [{\citenamefont {Gangloff}\ \emph {et~al.}(2019)\citenamefont
  {Gangloff}, \citenamefont {{\'E}thier-Majcher}, \citenamefont {Lang},
  \citenamefont {Denning}, \citenamefont {Bodey}, \citenamefont {Jackson},
  \citenamefont {Clarke}, \citenamefont {Hugues}, \citenamefont {Le~Gall},\
  and\ \citenamefont {Atat{\"u}re}}]{gangloff2019quantum}%
  \BibitemOpen
  \bibfield  {author} {\bibinfo {author} {\bibfnamefont {D.~A.}\ \bibnamefont
  {Gangloff}}, \bibinfo {author} {\bibfnamefont {G.}~\bibnamefont
  {{\'E}thier-Majcher}}, \bibinfo {author} {\bibfnamefont {C.}~\bibnamefont
  {Lang}}, \bibinfo {author} {\bibfnamefont {E.~V.}\ \bibnamefont {Denning}},
  \bibinfo {author} {\bibfnamefont {J.~H.}\ \bibnamefont {Bodey}}, \bibinfo
  {author} {\bibfnamefont {D.~M.}\ \bibnamefont {Jackson}}, \bibinfo {author}
  {\bibfnamefont {E.}~\bibnamefont {Clarke}}, \bibinfo {author} {\bibfnamefont
  {M.}~\bibnamefont {Hugues}}, \bibinfo {author} {\bibfnamefont
  {C.}~\bibnamefont {Le~Gall}},\ and\ \bibinfo {author} {\bibfnamefont
  {M.}~\bibnamefont {Atat{\"u}re}},\ }\href@noop {} {\bibfield  {journal}
  {\bibinfo  {journal} {Science}\ }\textbf {\bibinfo {volume} {364}},\ \bibinfo
  {pages} {62} (\bibinfo {year} {2019})}\BibitemShut {NoStop}%
\bibitem [{\citenamefont {Chesi}\ and\ \citenamefont
  {Coish}(2015)}]{chesi2015theory}%
  \BibitemOpen
  \bibfield  {author} {\bibinfo {author} {\bibfnamefont {S.}~\bibnamefont
  {Chesi}}\ and\ \bibinfo {author} {\bibfnamefont {W.~A.}\ \bibnamefont
  {Coish}},\ }\href@noop {} {\bibfield  {journal} {\bibinfo  {journal} {\prb}\
  }\textbf {\bibinfo {volume} {91}},\ \bibinfo {pages} {245306} (\bibinfo
  {year} {2015})}\BibitemShut {NoStop}%
\bibitem [{\citenamefont {Yang}\ \emph {et~al.}(2018)\citenamefont {Yang},
  \citenamefont {Willke}, \citenamefont {Bae}, \citenamefont {Ferr{\'o}n},
  \citenamefont {Lado}, \citenamefont {Ardavan}, \citenamefont
  {Fern{\'a}ndez-Rossier}, \citenamefont {Heinrich},\ and\ \citenamefont
  {Lutz}}]{yang2018electrically}%
  \BibitemOpen
  \bibfield  {author} {\bibinfo {author} {\bibfnamefont {K.}~\bibnamefont
  {Yang}}, \bibinfo {author} {\bibfnamefont {P.}~\bibnamefont {Willke}},
  \bibinfo {author} {\bibfnamefont {Y.}~\bibnamefont {Bae}}, \bibinfo {author}
  {\bibfnamefont {A.}~\bibnamefont {Ferr{\'o}n}}, \bibinfo {author}
  {\bibfnamefont {J.~L.}\ \bibnamefont {Lado}}, \bibinfo {author}
  {\bibfnamefont {A.}~\bibnamefont {Ardavan}}, \bibinfo {author} {\bibfnamefont
  {J.}~\bibnamefont {Fern{\'a}ndez-Rossier}}, \bibinfo {author} {\bibfnamefont
  {A.~J.}\ \bibnamefont {Heinrich}},\ and\ \bibinfo {author} {\bibfnamefont
  {C.~P.}\ \bibnamefont {Lutz}},\ }\href@noop {} {\bibfield  {journal}
  {\bibinfo  {journal} {Nat.~Nanotechnol.~}\ }\textbf {\bibinfo {volume}
  {13}},\ \bibinfo {pages} {1120} (\bibinfo {year} {2018})}\BibitemShut
  {NoStop}%
\bibitem [{\citenamefont {Kadowaki}\ and\ \citenamefont
  {Nishimori}(1998)}]{kadowaki1998quantum}%
  \BibitemOpen
  \bibfield  {author} {\bibinfo {author} {\bibfnamefont {T.}~\bibnamefont
  {Kadowaki}}\ and\ \bibinfo {author} {\bibfnamefont {H.}~\bibnamefont
  {Nishimori}},\ }\href@noop {} {\bibfield  {journal} {\bibinfo  {journal}
  {\pre}\ }\textbf {\bibinfo {volume} {58}},\ \bibinfo {pages} {5355} (\bibinfo
  {year} {1998})}\BibitemShut {NoStop}%
\bibitem [{\citenamefont {Pudenz}\ \emph {et~al.}(2014)\citenamefont {Pudenz},
  \citenamefont {Albash},\ and\ \citenamefont {Lidar}}]{pudenz2014error}%
  \BibitemOpen
  \bibfield  {author} {\bibinfo {author} {\bibfnamefont {K.~L.}\ \bibnamefont
  {Pudenz}}, \bibinfo {author} {\bibfnamefont {T.}~\bibnamefont {Albash}},\
  and\ \bibinfo {author} {\bibfnamefont {D.~A.}\ \bibnamefont {Lidar}},\
  }\href@noop {} {\bibfield  {journal} {\bibinfo  {journal} {Nat.~Commun.~}\
  }\textbf {\bibinfo {volume} {5}} (\bibinfo {year} {2014})}\BibitemShut
  {NoStop}%
\bibitem [{\citenamefont {Ozfidan}\ \emph {et~al.}(2019)\citenamefont
  {Ozfidan}, \citenamefont {Deng}, \citenamefont {Smirnov}, \citenamefont
  {Lanting}, \citenamefont {Harris}, \citenamefont {Swenson}, \citenamefont
  {Whittaker}, \citenamefont {Altomare}, \citenamefont {Babcock}, \citenamefont
  {Baron} \emph {et~al.}}]{ozfidan2019demonstration}%
  \BibitemOpen
  \bibfield  {author} {\bibinfo {author} {\bibfnamefont {I.}~\bibnamefont
  {Ozfidan}}, \bibinfo {author} {\bibfnamefont {C.}~\bibnamefont {Deng}},
  \bibinfo {author} {\bibfnamefont {A.~Y.}\ \bibnamefont {Smirnov}}, \bibinfo
  {author} {\bibfnamefont {T.}~\bibnamefont {Lanting}}, \bibinfo {author}
  {\bibfnamefont {R.}~\bibnamefont {Harris}}, \bibinfo {author} {\bibfnamefont
  {L.}~\bibnamefont {Swenson}}, \bibinfo {author} {\bibfnamefont
  {J.}~\bibnamefont {Whittaker}}, \bibinfo {author} {\bibfnamefont
  {F.}~\bibnamefont {Altomare}}, \bibinfo {author} {\bibfnamefont
  {M.}~\bibnamefont {Babcock}}, \bibinfo {author} {\bibfnamefont
  {C.}~\bibnamefont {Baron}}, \emph {et~al.},\ }\href@noop {} {\bibfield
  {journal} {\bibinfo  {journal} {arXiv:1903.06139}\ } (\bibinfo {year}
  {2019})}\BibitemShut {NoStop}%
\bibitem [{\citenamefont {Denchev}\ \emph {et~al.}(2016)\citenamefont
  {Denchev}, \citenamefont {Boixo}, \citenamefont {Isakov}, \citenamefont
  {Ding}, \citenamefont {Babbush}, \citenamefont {Smelyanskiy}, \citenamefont
  {Martinis},\ and\ \citenamefont {Neven}}]{denchev2016computational}%
  \BibitemOpen
  \bibfield  {author} {\bibinfo {author} {\bibfnamefont {V.~S.}\ \bibnamefont
  {Denchev}}, \bibinfo {author} {\bibfnamefont {S.}~\bibnamefont {Boixo}},
  \bibinfo {author} {\bibfnamefont {S.~V.}\ \bibnamefont {Isakov}}, \bibinfo
  {author} {\bibfnamefont {N.}~\bibnamefont {Ding}}, \bibinfo {author}
  {\bibfnamefont {R.}~\bibnamefont {Babbush}}, \bibinfo {author} {\bibfnamefont
  {V.}~\bibnamefont {Smelyanskiy}}, \bibinfo {author} {\bibfnamefont
  {J.}~\bibnamefont {Martinis}},\ and\ \bibinfo {author} {\bibfnamefont
  {H.}~\bibnamefont {Neven}},\ }\href
  {https://doi.org/10.1103/PhysRevX.6.031015} {\bibfield  {journal} {\bibinfo
  {journal} {Phys.~Rev.~X}\ }\textbf {\bibinfo {volume} {6}},\ \bibinfo {pages}
  {031015} (\bibinfo {year} {2016})}\BibitemShut {NoStop}%
\bibitem [{\citenamefont {Dattani}\ \emph {et~al.}(2019)\citenamefont
  {Dattani}, \citenamefont {Szalay},\ and\ \citenamefont
  {Chancellor}}]{dattani2019pegasus}%
  \BibitemOpen
  \bibfield  {author} {\bibinfo {author} {\bibfnamefont {N.}~\bibnamefont
  {Dattani}}, \bibinfo {author} {\bibfnamefont {S.}~\bibnamefont {Szalay}},\
  and\ \bibinfo {author} {\bibfnamefont {N.}~\bibnamefont {Chancellor}},\
  }\href@noop {} {\bibfield  {journal} {\bibinfo  {journal} {arXiv:1901.07636}\
  } (\bibinfo {year} {2019})}\BibitemShut {NoStop}%
\bibitem [{\citenamefont {Eto}\ \emph {et~al.}(2004)\citenamefont {Eto},
  \citenamefont {Ashiwa},\ and\ \citenamefont {Murata}}]{eto2004current}%
  \BibitemOpen
  \bibfield  {author} {\bibinfo {author} {\bibfnamefont {M.}~\bibnamefont
  {Eto}}, \bibinfo {author} {\bibfnamefont {T.}~\bibnamefont {Ashiwa}},\ and\
  \bibinfo {author} {\bibfnamefont {M.}~\bibnamefont {Murata}},\ }\href@noop {}
  {\bibfield  {journal} {\bibinfo  {journal} {J.~Phys.~Soc.~Jpn.~}\ }\textbf
  {\bibinfo {volume} {73}},\ \bibinfo {pages} {307} (\bibinfo {year}
  {2004})}\BibitemShut {NoStop}%
\bibitem [{\citenamefont {Schuetz}\ \emph {et~al.}(2012)\citenamefont
  {Schuetz}, \citenamefont {Kessler}, \citenamefont {Cirac},\ and\
  \citenamefont {Giedke}}]{schuetz2012superradiance}%
  \BibitemOpen
  \bibfield  {author} {\bibinfo {author} {\bibfnamefont {M.~J.~A.}\
  \bibnamefont {Schuetz}}, \bibinfo {author} {\bibfnamefont {E.~M.}\
  \bibnamefont {Kessler}}, \bibinfo {author} {\bibfnamefont {J.~I.}\
  \bibnamefont {Cirac}},\ and\ \bibinfo {author} {\bibfnamefont
  {G.}~\bibnamefont {Giedke}},\ }\href@noop {} {\bibfield  {journal} {\bibinfo
  {journal} {\prb}\ }\textbf {\bibinfo {volume} {86}},\ \bibinfo {pages}
  {085322} (\bibinfo {year} {2012})}\BibitemShut {NoStop}%
\bibitem [{\citenamefont {Gross}\ and\ \citenamefont
  {Haroche}(1982)}]{gross1982superradiance}%
  \BibitemOpen
  \bibfield  {author} {\bibinfo {author} {\bibfnamefont {M.}~\bibnamefont
  {Gross}}\ and\ \bibinfo {author} {\bibfnamefont {S.}~\bibnamefont
  {Haroche}},\ }\href@noop {} {\bibfield  {journal} {\bibinfo  {journal}
  {Phys.~Rep.~}\ }\textbf {\bibinfo {volume} {93}},\ \bibinfo {pages} {301}
  (\bibinfo {year} {1982})}\BibitemShut {NoStop}%
\bibitem [{\citenamefont {Morello}\ \emph {et~al.}(2010)\citenamefont
  {Morello}, \citenamefont {Pla}, \citenamefont {Zwanenburg}, \citenamefont
  {Chan}, \citenamefont {Tan}, \citenamefont {Huebl}, \citenamefont
  {M{\"o}tt{\"o}nen}, \citenamefont {Nugroho}, \citenamefont {Yang},
  \citenamefont {van Donkelaar} \emph {et~al.}}]{morello2010single}%
  \BibitemOpen
  \bibfield  {author} {\bibinfo {author} {\bibfnamefont {A.}~\bibnamefont
  {Morello}}, \bibinfo {author} {\bibfnamefont {J.~J.}\ \bibnamefont {Pla}},
  \bibinfo {author} {\bibfnamefont {F.~A.}\ \bibnamefont {Zwanenburg}},
  \bibinfo {author} {\bibfnamefont {K.~W.}\ \bibnamefont {Chan}}, \bibinfo
  {author} {\bibfnamefont {K.~Y.}\ \bibnamefont {Tan}}, \bibinfo {author}
  {\bibfnamefont {H.}~\bibnamefont {Huebl}}, \bibinfo {author} {\bibfnamefont
  {M.}~\bibnamefont {M{\"o}tt{\"o}nen}}, \bibinfo {author} {\bibfnamefont
  {C.~D.}\ \bibnamefont {Nugroho}}, \bibinfo {author} {\bibfnamefont
  {C.}~\bibnamefont {Yang}}, \bibinfo {author} {\bibfnamefont {J.~A.}\
  \bibnamefont {van Donkelaar}}, \emph {et~al.},\ }\href@noop {} {\bibfield
  {journal} {\bibinfo  {journal} {Nature}\ }\textbf {\bibinfo {volume} {467}},\
  \bibinfo {pages} {687} (\bibinfo {year} {2010})}\BibitemShut {NoStop}%
\bibitem [{\citenamefont {He}\ \emph {et~al.}(2019{\natexlab{b}})\citenamefont
  {He}, \citenamefont {Gorman}, \citenamefont {Keith}, \citenamefont {Kranz},
  \citenamefont {Keizer},\ and\ \citenamefont {Simmons}}]{he2019two}%
  \BibitemOpen
  \bibfield  {author} {\bibinfo {author} {\bibfnamefont {Y.}~\bibnamefont
  {He}}, \bibinfo {author} {\bibfnamefont {S.}~\bibnamefont {Gorman}}, \bibinfo
  {author} {\bibfnamefont {D.}~\bibnamefont {Keith}}, \bibinfo {author}
  {\bibfnamefont {L.}~\bibnamefont {Kranz}}, \bibinfo {author} {\bibfnamefont
  {J.}~\bibnamefont {Keizer}},\ and\ \bibinfo {author} {\bibfnamefont
  {M.}~\bibnamefont {Simmons}},\ }\href@noop {} {\bibfield  {journal} {\bibinfo
   {journal} {Nature}\ }\textbf {\bibinfo {volume} {571}},\ \bibinfo {pages}
  {371} (\bibinfo {year} {2019}{\natexlab{b}})}\BibitemShut {NoStop}%
\bibitem [{\citenamefont {Petta}\ \emph {et~al.}(2005)\citenamefont {Petta},
  \citenamefont {Johnson}, \citenamefont {Taylor}, \citenamefont {Laird},
  \citenamefont {Yacoby}, \citenamefont {Lukin}, \citenamefont {Marcus},
  \citenamefont {Hanson},\ and\ \citenamefont {Gossard}}]{petta2005coherent}%
  \BibitemOpen
  \bibfield  {author} {\bibinfo {author} {\bibfnamefont {J.~R.}\ \bibnamefont
  {Petta}}, \bibinfo {author} {\bibfnamefont {A.~C.}\ \bibnamefont {Johnson}},
  \bibinfo {author} {\bibfnamefont {J.~M.}\ \bibnamefont {Taylor}}, \bibinfo
  {author} {\bibfnamefont {E.~A.}\ \bibnamefont {Laird}}, \bibinfo {author}
  {\bibfnamefont {A.}~\bibnamefont {Yacoby}}, \bibinfo {author} {\bibfnamefont
  {M.~D.}\ \bibnamefont {Lukin}}, \bibinfo {author} {\bibfnamefont {C.~M.}\
  \bibnamefont {Marcus}}, \bibinfo {author} {\bibfnamefont {M.~P.}\
  \bibnamefont {Hanson}},\ and\ \bibinfo {author} {\bibfnamefont {A.~C.}\
  \bibnamefont {Gossard}},\ }\href@noop {} {\bibfield  {journal} {\bibinfo
  {journal} {Science}\ }\textbf {\bibinfo {volume} {309}},\ \bibinfo {pages}
  {2180} (\bibinfo {year} {2005})}\BibitemShut {NoStop}%
\bibitem [{\citenamefont {Veldhorst}\ \emph {et~al.}(2015)\citenamefont
  {Veldhorst}, \citenamefont {Yang}, \citenamefont {Hwang}, \citenamefont
  {Huang}, \citenamefont {Dehollain}, \citenamefont {Muhonen}, \citenamefont
  {Simmons}, \citenamefont {Laucht}, \citenamefont {Hudson}, \citenamefont
  {Itoh} \emph {et~al.}}]{veldhorst2015two}%
  \BibitemOpen
  \bibfield  {author} {\bibinfo {author} {\bibfnamefont {M.}~\bibnamefont
  {Veldhorst}}, \bibinfo {author} {\bibfnamefont {C.}~\bibnamefont {Yang}},
  \bibinfo {author} {\bibfnamefont {J.}~\bibnamefont {Hwang}}, \bibinfo
  {author} {\bibfnamefont {W.}~\bibnamefont {Huang}}, \bibinfo {author}
  {\bibfnamefont {J.}~\bibnamefont {Dehollain}}, \bibinfo {author}
  {\bibfnamefont {J.}~\bibnamefont {Muhonen}}, \bibinfo {author} {\bibfnamefont
  {S.}~\bibnamefont {Simmons}}, \bibinfo {author} {\bibfnamefont
  {A.}~\bibnamefont {Laucht}}, \bibinfo {author} {\bibfnamefont
  {F.}~\bibnamefont {Hudson}}, \bibinfo {author} {\bibfnamefont {K.~M.}\
  \bibnamefont {Itoh}}, \emph {et~al.},\ }\href@noop {} {\bibfield  {journal}
  {\bibinfo  {journal} {Nature}\ }\textbf {\bibinfo {volume} {526}},\ \bibinfo
  {pages} {410} (\bibinfo {year} {2015})}\BibitemShut {NoStop}%
\bibitem [{\citenamefont {Fang}\ \emph {et~al.}(2020)\citenamefont {Fang},
  \citenamefont {Wang}, \citenamefont {Fazio},\ and\ \citenamefont
  {Chesi}}]{fang2020superradiant}%
  \BibitemOpen
  \bibfield  {author} {\bibinfo {author} {\bibfnamefont {Y.-N.}\ \bibnamefont
  {Fang}}, \bibinfo {author} {\bibfnamefont {Y.-D.}\ \bibnamefont {Wang}},
  \bibinfo {author} {\bibfnamefont {R.}~\bibnamefont {Fazio}},\ and\ \bibinfo
  {author} {\bibfnamefont {S.}~\bibnamefont {Chesi}},\ }\href@noop {}
  {\bibfield  {journal} {\bibinfo  {journal} {arXiv preprint arXiv:2002.01219}\
  } (\bibinfo {year} {2020})}\BibitemShut {NoStop}%
\bibitem [{\citenamefont {Studenikin}\ \emph {et~al.}(2012)\citenamefont
  {Studenikin}, \citenamefont {Thorgrimson}, \citenamefont {Aers},
  \citenamefont {Kam}, \citenamefont {Zawadzki}, \citenamefont {Wasilewski},
  \citenamefont {Bogan},\ and\ \citenamefont
  {Sachrajda}}]{studenikin2012enhanced}%
  \BibitemOpen
  \bibfield  {author} {\bibinfo {author} {\bibfnamefont {S.~A.}\ \bibnamefont
  {Studenikin}}, \bibinfo {author} {\bibfnamefont {J.}~\bibnamefont
  {Thorgrimson}}, \bibinfo {author} {\bibfnamefont {G.~C.}\ \bibnamefont
  {Aers}}, \bibinfo {author} {\bibfnamefont {A.}~\bibnamefont {Kam}}, \bibinfo
  {author} {\bibfnamefont {P.}~\bibnamefont {Zawadzki}}, \bibinfo {author}
  {\bibfnamefont {Z.}~\bibnamefont {Wasilewski}}, \bibinfo {author}
  {\bibfnamefont {A.}~\bibnamefont {Bogan}},\ and\ \bibinfo {author}
  {\bibfnamefont {A.~S.}\ \bibnamefont {Sachrajda}},\ }\href@noop {} {\bibfield
   {journal} {\bibinfo  {journal} {\apl}\ }\textbf {\bibinfo {volume} {101}},\
  \bibinfo {pages} {233101} (\bibinfo {year} {2012})}\BibitemShut {NoStop}%
\bibitem [{\citenamefont {Yang}\ \emph {et~al.}(2014)\citenamefont {Yang},
  \citenamefont {Rossi}, \citenamefont {Lai}, \citenamefont {Leon},
  \citenamefont {Lim},\ and\ \citenamefont {Dzurak}}]{yang2014charge}%
  \BibitemOpen
  \bibfield  {author} {\bibinfo {author} {\bibfnamefont {C.~H.}\ \bibnamefont
  {Yang}}, \bibinfo {author} {\bibfnamefont {A.}~\bibnamefont {Rossi}},
  \bibinfo {author} {\bibfnamefont {N.~S.}\ \bibnamefont {Lai}}, \bibinfo
  {author} {\bibfnamefont {R.}~\bibnamefont {Leon}}, \bibinfo {author}
  {\bibfnamefont {W.~H.}\ \bibnamefont {Lim}},\ and\ \bibinfo {author}
  {\bibfnamefont {A.~S.}\ \bibnamefont {Dzurak}},\ }\href@noop {} {\bibfield
  {journal} {\bibinfo  {journal} {\apl}\ }\textbf {\bibinfo {volume} {105}},\
  \bibinfo {pages} {183505} (\bibinfo {year} {2014})}\BibitemShut {NoStop}%
\bibitem [{\citenamefont {Harvey-Collard}\ \emph {et~al.}(2018)\citenamefont
  {Harvey-Collard}, \citenamefont {D’Anjou}, \citenamefont {Rudolph},
  \citenamefont {Jacobson}, \citenamefont {Dominguez}, \citenamefont
  {Ten~Eyck}, \citenamefont {Wendt}, \citenamefont {Pluym}, \citenamefont
  {Lilly}, \citenamefont {Coish}, \citenamefont {Pioro-Ladri\`ere},\ and\
  \citenamefont {Carroll}}]{harvey2018high}%
  \BibitemOpen
  \bibfield  {author} {\bibinfo {author} {\bibfnamefont {P.}~\bibnamefont
  {Harvey-Collard}}, \bibinfo {author} {\bibfnamefont {B.}~\bibnamefont
  {D’Anjou}}, \bibinfo {author} {\bibfnamefont {M.}~\bibnamefont {Rudolph}},
  \bibinfo {author} {\bibfnamefont {N.~T.}\ \bibnamefont {Jacobson}}, \bibinfo
  {author} {\bibfnamefont {J.}~\bibnamefont {Dominguez}}, \bibinfo {author}
  {\bibfnamefont {G.~A.}\ \bibnamefont {Ten~Eyck}}, \bibinfo {author}
  {\bibfnamefont {J.~R.}\ \bibnamefont {Wendt}}, \bibinfo {author}
  {\bibfnamefont {T.}~\bibnamefont {Pluym}}, \bibinfo {author} {\bibfnamefont
  {M.~P.}\ \bibnamefont {Lilly}}, \bibinfo {author} {\bibfnamefont {W.~A.}\
  \bibnamefont {Coish}}, \bibinfo {author} {\bibfnamefont {M.}~\bibnamefont
  {Pioro-Ladri\`ere}},\ and\ \bibinfo {author} {\bibfnamefont {M.~S.}\
  \bibnamefont {Carroll}},\ }\href@noop {} {\bibfield  {journal} {\bibinfo
  {journal} {Phys.~Rev.~X}\ }\textbf {\bibinfo {volume} {8}},\ \bibinfo {pages}
  {021046} (\bibinfo {year} {2018})}\BibitemShut {NoStop}%
\bibitem [{\citenamefont {Qassemi}\ \emph {et~al.}(2009)\citenamefont
  {Qassemi}, \citenamefont {Coish},\ and\ \citenamefont
  {Wilhelm}}]{qassemi2009stationary}%
  \BibitemOpen
  \bibfield  {author} {\bibinfo {author} {\bibfnamefont {F.}~\bibnamefont
  {Qassemi}}, \bibinfo {author} {\bibfnamefont {W.~A.}\ \bibnamefont {Coish}},\
  and\ \bibinfo {author} {\bibfnamefont {F.~K.}\ \bibnamefont {Wilhelm}},\
  }\href@noop {} {\bibfield  {journal} {\bibinfo  {journal} {\prl}\ }\textbf
  {\bibinfo {volume} {102}},\ \bibinfo {pages} {176806} (\bibinfo {year}
  {2009})}\BibitemShut {NoStop}%
\bibitem [{\citenamefont {Lai}\ \emph {et~al.}(2011)\citenamefont {Lai},
  \citenamefont {Lim}, \citenamefont {Yang}, \citenamefont {Zwanenburg},
  \citenamefont {Coish}, \citenamefont {Qassemi}, \citenamefont {Morello},\
  and\ \citenamefont {Dzurak}}]{lai2011pauli}%
  \BibitemOpen
  \bibfield  {author} {\bibinfo {author} {\bibfnamefont {N.}~\bibnamefont
  {Lai}}, \bibinfo {author} {\bibfnamefont {W.}~\bibnamefont {Lim}}, \bibinfo
  {author} {\bibfnamefont {C.}~\bibnamefont {Yang}}, \bibinfo {author}
  {\bibfnamefont {F.}~\bibnamefont {Zwanenburg}}, \bibinfo {author}
  {\bibfnamefont {W.~A.}\ \bibnamefont {Coish}}, \bibinfo {author}
  {\bibfnamefont {F.}~\bibnamefont {Qassemi}}, \bibinfo {author} {\bibfnamefont
  {A.}~\bibnamefont {Morello}},\ and\ \bibinfo {author} {\bibfnamefont
  {A.}~\bibnamefont {Dzurak}},\ }\href@noop {} {\bibfield  {journal} {\bibinfo
  {journal} {Sci.~Rep.~}\ }\textbf {\bibinfo {volume} {1}},\ \bibinfo {pages}
  {110} (\bibinfo {year} {2011})}\BibitemShut {NoStop}%
\bibitem [{\citenamefont {Maune}\ \emph {et~al.}(2012)\citenamefont {Maune},
  \citenamefont {Borselli}, \citenamefont {Huang}, \citenamefont {Ladd},
  \citenamefont {Deelman}, \citenamefont {Holabird}, \citenamefont {Kiselev},
  \citenamefont {Alvarado-Rodriguez}, \citenamefont {Ross}, \citenamefont
  {Schmitz} \emph {et~al.}}]{maune2012coherent}%
  \BibitemOpen
  \bibfield  {author} {\bibinfo {author} {\bibfnamefont {B.~M.}\ \bibnamefont
  {Maune}}, \bibinfo {author} {\bibfnamefont {M.~G.}\ \bibnamefont {Borselli}},
  \bibinfo {author} {\bibfnamefont {B.}~\bibnamefont {Huang}}, \bibinfo
  {author} {\bibfnamefont {T.~D.}\ \bibnamefont {Ladd}}, \bibinfo {author}
  {\bibfnamefont {P.~W.}\ \bibnamefont {Deelman}}, \bibinfo {author}
  {\bibfnamefont {K.~S.}\ \bibnamefont {Holabird}}, \bibinfo {author}
  {\bibfnamefont {A.~A.}\ \bibnamefont {Kiselev}}, \bibinfo {author}
  {\bibfnamefont {I.}~\bibnamefont {Alvarado-Rodriguez}}, \bibinfo {author}
  {\bibfnamefont {R.~S.}\ \bibnamefont {Ross}}, \bibinfo {author}
  {\bibfnamefont {A.~E.}\ \bibnamefont {Schmitz}}, \emph {et~al.},\ }\href@noop
  {} {\bibfield  {journal} {\bibinfo  {journal} {Nature}\ }\textbf {\bibinfo
  {volume} {481}},\ \bibinfo {pages} {344} (\bibinfo {year}
  {2012})}\BibitemShut {NoStop}%
\bibitem [{\citenamefont {Veldhorst}\ \emph {et~al.}(2014)\citenamefont
  {Veldhorst}, \citenamefont {Hwang}, \citenamefont {Yang}, \citenamefont
  {Leenstra}, \citenamefont {de~Ronde}, \citenamefont {Dehollain},
  \citenamefont {Muhonen}, \citenamefont {Hudson}, \citenamefont {Itoh},
  \citenamefont {Morello} \emph {et~al.}}]{veldhorst2014addressable}%
  \BibitemOpen
  \bibfield  {author} {\bibinfo {author} {\bibfnamefont {M.}~\bibnamefont
  {Veldhorst}}, \bibinfo {author} {\bibfnamefont {J.}~\bibnamefont {Hwang}},
  \bibinfo {author} {\bibfnamefont {C.}~\bibnamefont {Yang}}, \bibinfo {author}
  {\bibfnamefont {A.}~\bibnamefont {Leenstra}}, \bibinfo {author}
  {\bibfnamefont {B.}~\bibnamefont {de~Ronde}}, \bibinfo {author}
  {\bibfnamefont {J.}~\bibnamefont {Dehollain}}, \bibinfo {author}
  {\bibfnamefont {J.}~\bibnamefont {Muhonen}}, \bibinfo {author} {\bibfnamefont
  {F.}~\bibnamefont {Hudson}}, \bibinfo {author} {\bibfnamefont {K.~M.}\
  \bibnamefont {Itoh}}, \bibinfo {author} {\bibfnamefont {A.}~\bibnamefont
  {Morello}}, \emph {et~al.},\ }\href@noop {} {\bibfield  {journal} {\bibinfo
  {journal} {Nat.~Nanotechnol.~}\ }\textbf {\bibinfo {volume} {9}},\ \bibinfo
  {pages} {981} (\bibinfo {year} {2014})}\BibitemShut {NoStop}%
\bibitem [{\citenamefont {Asaad}\ \emph {et~al.}(2020)\citenamefont {Asaad},
  \citenamefont {Mourik}, \citenamefont {Joecker}, \citenamefont {Johnson},
  \citenamefont {Baczewski}, \citenamefont {Firgau}, \citenamefont
  {M{\k{a}}dzik}, \citenamefont {Schmitt}, \citenamefont {Pla}, \citenamefont
  {Hudson} \emph {et~al.}}]{asaad2020coherent}%
  \BibitemOpen
  \bibfield  {author} {\bibinfo {author} {\bibfnamefont {S.}~\bibnamefont
  {Asaad}}, \bibinfo {author} {\bibfnamefont {V.}~\bibnamefont {Mourik}},
  \bibinfo {author} {\bibfnamefont {B.}~\bibnamefont {Joecker}}, \bibinfo
  {author} {\bibfnamefont {M.~A.}\ \bibnamefont {Johnson}}, \bibinfo {author}
  {\bibfnamefont {A.~D.}\ \bibnamefont {Baczewski}}, \bibinfo {author}
  {\bibfnamefont {H.~R.}\ \bibnamefont {Firgau}}, \bibinfo {author}
  {\bibfnamefont {M.~T.}\ \bibnamefont {M{\k{a}}dzik}}, \bibinfo {author}
  {\bibfnamefont {V.}~\bibnamefont {Schmitt}}, \bibinfo {author} {\bibfnamefont
  {J.~J.}\ \bibnamefont {Pla}}, \bibinfo {author} {\bibfnamefont {F.~E.}\
  \bibnamefont {Hudson}}, \emph {et~al.},\ }\href@noop {} {\bibfield  {journal}
  {\bibinfo  {journal} {Nature}\ }\textbf {\bibinfo {volume} {579}},\ \bibinfo
  {pages} {205} (\bibinfo {year} {2020})}\BibitemShut {NoStop}%
\bibitem [{\citenamefont {Johnson}\ \emph {et~al.}(2011)\citenamefont
  {Johnson}, \citenamefont {Amin}, \citenamefont {Gildert}, \citenamefont
  {Lanting}, \citenamefont {Hamze}, \citenamefont {Dickson}, \citenamefont
  {Harris}, \citenamefont {Berkley}, \citenamefont {Johansson}, \citenamefont
  {Bunyk} \emph {et~al.}}]{johnson2011quantum}%
  \BibitemOpen
  \bibfield  {author} {\bibinfo {author} {\bibfnamefont {M.~W.}\ \bibnamefont
  {Johnson}}, \bibinfo {author} {\bibfnamefont {M.~H.}\ \bibnamefont {Amin}},
  \bibinfo {author} {\bibfnamefont {S.}~\bibnamefont {Gildert}}, \bibinfo
  {author} {\bibfnamefont {T.}~\bibnamefont {Lanting}}, \bibinfo {author}
  {\bibfnamefont {F.}~\bibnamefont {Hamze}}, \bibinfo {author} {\bibfnamefont
  {N.}~\bibnamefont {Dickson}}, \bibinfo {author} {\bibfnamefont
  {R.}~\bibnamefont {Harris}}, \bibinfo {author} {\bibfnamefont {A.~J.}\
  \bibnamefont {Berkley}}, \bibinfo {author} {\bibfnamefont {J.}~\bibnamefont
  {Johansson}}, \bibinfo {author} {\bibfnamefont {P.}~\bibnamefont {Bunyk}},
  \emph {et~al.},\ }\href@noop {} {\bibfield  {journal} {\bibinfo  {journal}
  {Nature}\ }\textbf {\bibinfo {volume} {473}},\ \bibinfo {pages} {194}
  (\bibinfo {year} {2011})}\BibitemShut {NoStop}%
\bibitem [{\citenamefont {Matsuura}\ \emph {et~al.}(2016)\citenamefont
  {Matsuura}, \citenamefont {Nishimori}, \citenamefont {Albash},\ and\
  \citenamefont {Lidar}}]{matsuura2016mean}%
  \BibitemOpen
  \bibfield  {author} {\bibinfo {author} {\bibfnamefont {S.}~\bibnamefont
  {Matsuura}}, \bibinfo {author} {\bibfnamefont {H.}~\bibnamefont {Nishimori}},
  \bibinfo {author} {\bibfnamefont {T.}~\bibnamefont {Albash}},\ and\ \bibinfo
  {author} {\bibfnamefont {D.~A.}\ \bibnamefont {Lidar}},\ }\href@noop {}
  {\bibfield  {journal} {\bibinfo  {journal} {\prl}\ }\textbf {\bibinfo
  {volume} {116}},\ \bibinfo {pages} {220501} (\bibinfo {year}
  {2016})}\BibitemShut {NoStop}%
\bibitem [{\citenamefont {Matsuura}\ \emph {et~al.}(2017)\citenamefont
  {Matsuura}, \citenamefont {Nishimori}, \citenamefont {Vinci}, \citenamefont
  {Albash},\ and\ \citenamefont {Lidar}}]{matsuura2017quantum}%
  \BibitemOpen
  \bibfield  {author} {\bibinfo {author} {\bibfnamefont {S.}~\bibnamefont
  {Matsuura}}, \bibinfo {author} {\bibfnamefont {H.}~\bibnamefont {Nishimori}},
  \bibinfo {author} {\bibfnamefont {W.}~\bibnamefont {Vinci}}, \bibinfo
  {author} {\bibfnamefont {T.}~\bibnamefont {Albash}},\ and\ \bibinfo {author}
  {\bibfnamefont {D.~A.}\ \bibnamefont {Lidar}},\ }\href@noop {} {\bibfield
  {journal} {\bibinfo  {journal} {\pra}\ }\textbf {\bibinfo {volume} {95}},\
  \bibinfo {pages} {022308} (\bibinfo {year} {2017})}\BibitemShut {NoStop}%
\bibitem [{\citenamefont {Fick}\ and\ \citenamefont
  {Sauermann}(1990)}]{fick1990quantum}%
  \BibitemOpen
  \bibfield  {author} {\bibinfo {author} {\bibfnamefont {E.}~\bibnamefont
  {Fick}}\ and\ \bibinfo {author} {\bibfnamefont {G.}~\bibnamefont
  {Sauermann}},\ }\href@noop {} {\emph {\bibinfo {title} {The quantum
  statistics of dynamic processes}}},\ Vol.~\bibinfo {volume} {86}\ (\bibinfo
  {publisher} {Springer},\ \bibinfo {year} {1990})\BibitemShut {NoStop}%
\end{thebibliography}%
\end{document}